\documentclass[12pt]{article}
\usepackage{amsmath,amssymb,exscale,cancel}
\usepackage{slashed}
\usepackage{graphicx}
\usepackage{epsfig}
\usepackage{multicol}
\usepackage{color}
\usepackage{mathrsfs}
\usepackage{blindtext}
 \usepackage{fancyhdr}
\usepackage{hyperref}
\usepackage{cite}
\usepackage{mathtools}

\usepackage{floatrow}

\usepackage{graphicx}
\usepackage{sidecap}

\usepackage[latin1]{inputenc} 

\usepackage{multicol}  
 
 \usepackage{titlesec}
 
 \usepackage{rotating,slashed,xcolor,amsfonts,expdlist,charter}

\numberwithin{equation}{section}

\usepackage{xcolor}
\usepackage{sectsty}

\usepackage{mdframed}
\usepackage{titletoc}

\numberwithin{equation}{section}

\usepackage{xcolor}
\usepackage{sectsty}

\usepackage{mdframed}
\usepackage{titletoc}

\definecolor{secnum}{RGB}{13,151,225}
\definecolor{ptcbackground}{RGB}{212,237,252}
\definecolor{ptctitle}{RGB}{0,177,235}

\titlecontents{lsection}
  [5.8em]{\sffamily}
  {\color{secnum}\contentslabel{2.3em}\normalcolor}{}
  {\titlerule*[1000pc]{.}\contentspage\\\hspace*{-5.8em}\vspace*{5pt}%
    \color{white}\rule{\dimexpr\textwidth-15.5pt\relax}{1pt}}

\usepackage{hyperref}
\hypersetup{colorlinks,bookmarksopen,bookmarksnumbered,citecolor=rossos,
linkcolor=redy,pdfstartview=FitH,urlcolor=green-go}
\usepackage{slashed}

\definecolor{blus}{cmyk}{1,0.9,0,0.1}
\definecolor{verdes}{cmyk}{0.99,0,0.59,0.65}
\definecolor{rossos}{cmyk}{0,1,1,0.55}
\definecolor{redy}{cmyk}{0,1,1,0.7}
\definecolor{greeny}{cmyk}{0.99,0,0.59,0.98}
\definecolor{green-go}{cmyk}{0.79,0,0.59,0.5}

\usepackage{titlesec}

\newcommand{\beq}{\begin{equation}}
\newcommand{\eeq}{\end{equation}}

\def\hhref#1{\href{http://arxiv.org/abs/#1}{arXiv:#1}} 

\newcommand{\tmtextbf}[1]{{\bfseries{#1}}}
\newcommand{\tmtextrm}[1]{{\rmfamily{#1}}}
\newcommand{\bp}{\bar M_P}

\def\be{\begin{equation}}
\def\ee{\end{equation}}
\def\ba{\begin{array} }

\def\bac{\begin{array} {c}}
\def\bacc{\begin{array} {cc}}
\def\baccc{\begin{array} {ccc}}
\def\bacccc{\begin{array} {cccc}}
\def\ea{\end{array}}
\def\bea{\begin{eqnarray}}
\def\eea{\end{eqnarray}}

\definecolor{red}{rgb}{1,0,0}

\def\psl{\hbox{\hbox{${p}$}}\kern-1.9mm{\hbox{${/}$}}}
\def\dsl{\hbox{\hbox{${\partial}$}}\kern-2.2mm{\hbox{${/}$}}}
\def\Dsl{\hbox{\hbox{${D}$}}\kern-2.6mm{\hbox{${/}$}}}

\newcommand{\gappeq}{{\rlap{{\raise}.5ex\text{\ensuremath{>}}}{{\lower}.5ex\text{\ensuremath{\sim}}}}}
\newcommand{\lappeq}{{\rlap{{\raise}.5ex\text{\ensuremath{<}}}{{\lower}.5ex\text{\ensuremath{\sim}}}}}
\newcommand{\I}{\tmtextrm{1{\kern}-.24em l}}

\begin{document}
\topmargin -1.0cm
\oddsidemargin 0.9cm
\evensidemargin -0.5cm

{\vspace{-1cm}}
\begin{center}

\vspace{-1cm}

 {\LARGE \tmtextbf{ 
\color{rossos} \hspace{-0.9cm}    \\
Primordial Black Holes (as Dark Matter)  from the Supercooled Phase Transitions 
with \\ \vspace{.1cm} Radiative Symmetry Breaking
 \hspace{-1.6cm}}} {\vspace{.5cm}}\\



\vspace{1.3cm}

{\large{\bf   Indra Kumar Banerjee$^{a}$, Francesco Rescigno$^{b,c}$, Alberto Salvio$^{b,c}$}}

{\em  
\vspace{.4cm}

$^a$  Department of Physical Sciences, Indian Institute of Science Education and Research Berhampur,
Transit Campus, Government ITI, Berhampur 760010, Odisha, India

\vspace{0.6cm}

$^b$ Physics Department, University of Rome Tor Vergata, \\ 
via della Ricerca Scientifica, I-00133 Rome, Italy\\

\vspace{0.6cm}

$^c$ I. N. F. N. -  Rome Tor Vergata,\\
via della Ricerca Scientifica, I-00133 Rome, Italy\\

%
%
%
%
%
%
%
 \vspace{0.5cm}

%
%
%
%
%
}
%
%
\end{center}

%
%
\noindent ---------------------------------------------------------------------------------------------------------------------------------
\begin{center}
{\bf \large Abstract}
\end{center}
\noindent   We study in detail the production of primordial black holes (PBHs), as well as their mass and initial spin, due to the phase transitions corresponding to  radiative symmetry breaking (RSB) and featuring a large supercooling. The latter property allows us to use a model-independent approach. In this context, we demonstrate that the decay rate of the false vacuum grows exponentially with time to a high degree of accuracy, justifying a time dependence commonly assumed in the literature.
Our study provides ready-to-use results for determining the abundance, mass and initial spin of PBHs generated in a generic RSB model with large supercooling. We find that PBHs are generically produced in a broad region of the model-independent parameter space.  As an application, we  identify the subregion that may explain  recently reported microlensing anomalies. Additionally, we show that a simple Standard-Model extension, with right-handed neutrinos and  gauged $B-L$ featuring RSB, may explain an anomaly of this sort in a region of its parameter space.

\vspace{0.7cm}

\noindent---------------------------------------------------------------------------------------------------------------------------------

  \vspace{-0.9cm}
  
  \newpage 
\tableofcontents

\noindent

\vspace{0.5cm}

\section{Introduction}\label{Introduction}

The nature of dark matter (DM) is one of the greatest mysteries in physics and a key observational signals of new physics beyond the Standard Model (SM) of elementary particles.
The role of DM can be played by an elementary particle or by a condensate of some (or even many) elementary particles.  

Regarding the latter case, in the last years there has been a renewed interest in the possibility that DM might originate from   PBHs, black holes formed in the early universe. At the fundamental level, indeed, black holes, like all extended objects, must be a condensate of a (generically huge) number of elementary particles. The interest in the physics of compact objects has been fuelled by the discovery of gravitational waves (GWs) from binary black hole and neutron star mergers~\cite{Abbott:2016blz,TheLIGOScientific:2016wyq1,LIGOScientific:2017ync}.

In many realizations,  achieving  sufficiently large PBH production leads to ad hoc model building and/or significant parameter fine-tuning. This is, for instance, the case in single-field inflationary models~\cite{Cole:2023wyx}. Besides inflation, however, there are other mechanisms for producing  PBHs. One notable example is phase transitions (PTs), specifically  first-order phase transitions (FOPTs). FOPTs  are known to produce a sufficient amount of PBHs to account for  a fraction or even the entire DM abundance~\cite{Hawking:1982ga,Crawford:1982yz,Kodama:1982sf,Moss:1994iq,Khlopov:1999ys,Khlopov:2008qy,Gross:2021qgx,Kawana:2021tde,Liu:2021svg,Baker:2021sno,Jung:2021mku,Hashino:2021qoq,Hashino:2022tcs,Kawana:2022olo,Lewicki:2023ioy,Gouttenoire:2023naa,Gouttenoire:2023bqy,Salvio:2023ext,Gouttenoire:2023pxh,Salvio:2023blb,Lewicki:2024ghw,Balaji:2024rvo,Cai:2024nln,Ai:2024cka,Arteaga:2024vde,Goncalves:2024vkj,Banerjee:2024fam}. FOPTs are interesting also because they leave other potentially observable imprints, such as GWs, and they can be evidence for new physics, as the SM does not feature such transitions. 

In general terms, it is not possible to perturbatively describe FOPTs in models where the corresponding symmetry breaking arises entirely from the Higgs mechanism, because of the presence of complex effective potentials~\cite{Salvio:2020axm,Salvio:2023qgb,Salvio:2024upo} and other issues at high-temperatures~\cite{Weinberg:1974hy,Linde:1980ts,Linde:1978px}. These problems can be avoided in theories where  the corresponding symmetry breaking is mostly induced (and masses are generated) radiatively, i.e.~through perturbative loop corrections\footnote{More generally, it is not necessary for the entire theory to rely on this mechanism; it suffices for a sector of the theory with sufficiently small couplings to the rest to exhibit this behavior.}~\cite{Salvio:2023qgb,Salvio:2024upo}. Moreover, when this radiative symmetry breaking (RSB) occurs the corresponding PT is always of first order~\cite{Salvio:2023qgb,Salvio:2024upo}. The seminal work on RSB is Ref.~\cite{Coleman:1973jx} by Coleman and E.~Weinberg, who studied a simple toy model (see also Ref.~\cite{Levi:2022bzt}
for a subsequent analysis). The Coleman-Weinberg work was then extended to broader field theories by Gildener and S.~Weinberg~\cite{Gildener:1976ih}. Furthermore, thanks to an approximate scale invariance, in the RSB scenario the FOPTs feature a period of supercooling, when the temperature drops by several orders of magnitude below the critical temperature~\cite{Witten:1980ez,Salvio:2023qgb}.
Supercooling is one of the key properties of the RSB scenario that validate the one-loop perturbative approximation (as well as the derivative expansion)~\cite{Salvio:2023qgb}.

In~\cite{Salvio:2023qgb,Salvio:2023ext} it  was shown that a model-independent description of PTs in the RSB scenario is possible in terms of few parameters (which are computable once the model is specified) if enough supercooling occurs; this leads to an expansion in terms of a  quantity that is small when supercooling is large enough. 

The objective of the present work is to use such model-independent approach to explore the generality of the PBH production mechanism in perturbative FOPTs, without the need to specify a particular model. Since the approach is model-independent, this helps identify  not only possible parameter fine-tuning, but also possible fine-tuning in model space. Another motivation for this general analysis is the fact that can lead to the identification of the parameter space where the produced PBHs can have observable effects and even account for DM; this can be very useful to determine if and when a given model predicts PBH DM without repeating the analysis individually. To maintain model independence, we focus here on a PBH production mechanism that {\it always} operates in RSB theories with large-enough supercooling. Such a mechanism  
has been studied in a number of papers (see e.g.~Refs.~\cite{Kodama:1982sf,Liu:2021svg,Hashino:2021qoq,Kawana:2022olo,Gouttenoire:2023naa}) and exploits the non-zero probability (which is sizable if supercooling is strong) that some patches of the universe undergo RSB later than  the average of the other patches (here called the background). We refer to this  as the late blooming mechanism. While other mechanisms for PBH production and alternative DM candidates could contribute to the total DM abundance in specific models, these require analysis on a case-by-case basis within the RSB framework.

One of the key assumptions that, so far, has  often been made in PBH production through supercooled PTs is that the (false-vacuum) decay rate is an exponentially growing function of time,  with a time constant given by the logarithmic time derivative of the decay rate evaluated at the nucleation time, $\beta$. Another purpose of this work is to fill this loophole and justify this approximation in the class of perturbative theories featuring supercooled FOPTs\footnote{Ref.~\cite{Kanemura:2024pae} pointed out that this is a challenging task working in a model where symmetry breaking arises entirely from the Higgs mechanism.}. Once this justification is given, it is possible to understand that the smaller $\beta/H_n$ is the more efficient the late-blooming mechanism turns out to be, where $H_n$ is the Hubble rate at the nucleation time.

Generically, $\beta$ is one of the key parameters in FOPTs, so here  we also aim at studying  the $\beta$ dependence  of the scale factors, Hubble rates and energy densities for the background and the late-blooming patch.

The paper is structured as follows. In Sec.~\ref{A fresh look at the late-blooming mechanism} (and in the appendix) we will provide a fresh look at the late-blooming mechanism in general supercooled FOPTs. Sec.~\ref{General supercooled phase transitions from RSB and PBHs} will be then devoted to the study of this PBH production mechanism in the general framework of perturbative theories featuring supercooled FOPTs, with RSB.  There, in Subsec.~\ref{Time dependence of the false-vacuum decay rate} we will address the issue of the time dependence of the decay rate. Also, in Subsec.~\ref{PBH abundance, mass and initial PBH spin} we will calculate, through the large-supercooling model-independent approach, the PBH abundance, mass and initial  spin, varying the model-independent parameters. Sec.~\ref{Conclusions} provides a detailed summary of the main results  and the final conclusions.

\section{A fresh look at a PBH-production mechanism}\label{A fresh look at the late-blooming mechanism}

In this section we re-examine the late-blooming mechanism for PBH production in a FOPTs (see e.g. Refs.~\cite{Kodama:1982sf,Liu:2021svg,Hashino:2021qoq,Kawana:2022olo,Gouttenoire:2023naa,Cai:2024nln} for an introduction). The basic idea behind it is strongly linked to the probabilistic nature of the false-vacuum decay in a FOPT: there is a non zero probability (of thermal and/or quantum nature) that some patches of the universe remain in the false vacuum for a longer time than the background (i.e.~than the average time of the other patches); this creates a sufficiently-high mass excess
between the late patch (LP) and the background, such that the overdense region collapses to a black hole.

We will assume that the universe is  filled with radiation just before the FOPT occurs. The decay of the metastable vacuum is energetically favorable from the time $t_c$ where the temperature $T$ goes below its critical value $T_c$. After $t_c$,  true vacuum bubbles may start to appear and at a certain time $t_{\rm eq}$ the temperature reaches a value  $T_{\rm eq}$ at which the radiation energy density $\rho_R$ equals the difference $\Delta V$ between the false-vacuum and true-vacuum energy densities\footnote{In this work we use units where the Boltzmann constant, the speed of light and  the reduced Planck constant  are all equal to one.}:
\begin{equation}
	\label{eq:DV}
	\frac{\pi^{2}}{30}g_{*}(T_{\rm eq})T^4_{\rm eq}\equiv \Delta V ,
\end{equation}
where $g_*(T)$ is the effective number of relativistic species at temperature $T$. 

The (finite-temperature) decay rate per unit of three-dimensional volume of the false vacuum~\cite{Coleman:1977py,Callan:1977pt,Linde:1980tt,Linde:1981zj}, $\Gamma$, can be expanded for a generic time $t$ around the nucleation time $t_n$, when the PT starts to be effective, defined here as  the time  when   $\Gamma(t_n)=H(t_n)^4$, also\footnote{As usual a dot denotes a time derivative.} $H=\dot a/a$ is the Hubble rate and $a$ is the scale factor:
\begin{equation}
	\label{eq:GammaGen}
    \Gamma(t) = \Gamma(t_n)\text{exp}(\beta(t-t_n)+ \beta_2(t-t_n)^2+ \beta_3(t-t_n)^3+...).
\end{equation}
The parameter $\beta$ is defined by
\begin{equation}
	\beta \equiv \frac{1}{\Gamma(t)}\frac{d\Gamma(t)}{dt}\bigg|_{t=t_n}
\end{equation}
 and has been interpreted as the inverse duration of the PT in models with large-enough supercooling~\cite{Caprini:2015zlo,Caprini:2018mtu,vonHarling:2019gme}. 
Neglecting the coefficients ($\beta_2, \beta_3, ...$) of the higher powers of $t-t_n$  leads to
 \begin{equation}
	\label{eq:Gamma}
    \Gamma(t) \approx H_n^4 e^{\beta(t-t_n)}.
\end{equation}
where $ H_n\equiv H(t_n)$.
In Sec.~\ref{General supercooled phase transitions from RSB and PBHs} we will show that this approximation is valid in the RSB PTs, which feature a large amount of supercooling.

The strength of the FOPT is defined as usual by
\begin{equation}
	\label{eq:alpha}
	\alpha\equiv\left(\frac{\Delta V}{\rho_R} \right)_{T=T_n}=\left(\frac{T_{\text{eq}}}{T_n}\right)^4,
\end{equation}
where   $T_{n}$ is the nucleation temperature, the temperature at $t=t_n$.

Following~\cite{Gouttenoire:2023naa} we will assume that the full energy density $\rho$ is given by the sum of $\rho_R$ and the  vacuum energy density $\rho_V$:
\begin{equation}
	\rho=\rho_V + \rho_R.
\end{equation}
The formation of a bubble of true vacuum inside a false-vacuum environment is a probabilistic process so there is a non zero probability that in the $i$th patch of the universe
no bubble forms until a time $t_{n_i}>t_n$. In this case, at some point the radiation energy density outside such LP is smaller than the radiation energy density inside the LP (radiation energy density decreases with time). So, the LP eventually collapses if the mass excess
between the inside and the outside of the LP is large enough, which occurs if $t_{n_i}$ is sufficiently larger than $t_n$.
To study the probability that a PBH forms via such late-blooming mechanism and estimate the PBH abundance, we will model this system through a LP and some overall (average) background. 
The energy densities $\rho_R$ and $\rho_V$ will depend on the time $t$ and on the instant $t_{n_i}$ when the first bubble forms: for the background we will set $t_{n_i}= t_c$, while for the LP we choose $t_{n_i} = t_{n_i}^{ \rm PBH}$, where $t_{n_i}^{ \rm PBH}$ is the smallest time that allows the mass excess $\delta$ (that will be precisely defined below) to reach the critical value, $\delta_c$. We will approximate the various energy densities associated with a patch with  their
  space averages in the given patch, such that these quantities are homogeneous by construction.  In the numerical calculation we will use the benchmark 
value\footnote{This value for the mass excess in the Hubble-size LP has been calculated in the context of cosmological PBHs, see e.g.~\cite{Musco_2005, Polnarev_2007, Musco_2019, Musco_2021}.} $\delta_c = 0.45$, unless otherwise stated.

Since $\Gamma$ increases exponentially with $\beta$ at $t>t_n$ if~(\ref{eq:Gamma}) holds (which is the case in RSB PTs, see Sec.~\ref{General supercooled phase transitions from RSB and PBHs}) smaller values of $\beta$ lead to a longer period of vacuum domination and to the possibility of  a larger difference between $t_{n_i}$ and $t_n$. This explains why smaller values of $\beta$ corresponds to a larger PBH production and, as a result, why a supercooled PT can lead to a significant amount of PBHs.

For the background we have:
\begin{equation}
	\label{eq:BKG}
	\begin{cases}
		\dot{\rho}_R^{b}(t)+4\frac{\dot{a}_{b}(t)}{a_{b}(t)}\rho_R^{b}(t)=-\dot{\rho}^b_V(t),\\
		\frac{\dot{a}_{b}(t)}{a_{b}(t)}=\sqrt{\frac{\rho_V^{b}(t)+\rho_R^{b}(t)}{3\bp^2}},
	\end{cases},
\end{equation}
while for the LP
\begin{equation}
	\label{eq:LHP}
	\begin{cases}
		\dot{\rho}_R^{l}(t)+4\frac{\dot{a}_{l}(t)}{a_{l}(t)}\rho_R^{l}(t)=-\dot{\rho}^l_V(t),\\
		\frac{\dot{a}_{l}(t)}{a_{l}(t)}=\sqrt{\frac{\rho_V^{l}(t)+\rho_R^{l}(t)}{3\bp^2}},
	\end{cases},
\end{equation}
where $\bp$ is the reduced Planck mass and we understood the dependence on $t_{n_i}$. Also, we labeled the functions corresponding to  the background and the LP with $b$ and $l$, respectively. This notation differs from that of~\cite{Gouttenoire:2023naa,Liu:2021svg} and transparently shows that there are two Hubble rates, one for the observer in the background, $H_b = \dot a_b/a_b$, and one for the observer in the LP, $H_l= \dot a_l/a_l$:  in fact we expect that a period of exponential expansion first occurs in the whole universe and then continues only in the LP (until it eventually stops even there thanks, as always,  to vacuum decay).  This is possible only if  two observers inside and outside the LP measure different Hubble rates.

  We can write the time-dependent vacuum energy density $\rho_V(t)$
  after bubbles start to form as 
\begin{equation}
	\label{eq:FRHO}
	\rho_V(t) = \Delta V F(t,t_{n_i}).
\end{equation} 
 The time-dependent factor  $F(t,t_{n_i})$, which satisfies the condition $F(t_{n_i},t_{n_i})=1$, has been previously determined and discussed in~\cite{Guth:1979bh,Turner:1992tz} (see also~\cite{Gouttenoire:2023naa}):
\begin{equation}
	\label{eq:F}
	F(t,t_{n_i})= \text{exp}\left(-\int^{t}_{t_{n_i}}dt'\Gamma(t')a(t')^3\frac{4}{3}\pi r^3(t,t')\right)\equiv e^{-I(t,t_{n_i})},
\end{equation} 
where 
\be r(t,t')=\int_{t'}^t\frac{d\tilde{t}}{a(\tilde{t})}\ee
 is the comoving radius at time $t$ of a bubble that formed at time $t'$, assuming that the bubble walls propagates at the speed of light. This is the case in supercooled PTs, such as those associated with RSB, because the long-lasting exponential expansion that occurs in these PTs dilutes preexisting matter and radiation\footnote{When the bubble walls propagate at speed $v_w$ one has to replace $r(t,t')=\int_{t'}^td\tilde{t}\, v_w(\tilde{t})/a(\tilde{t})$.}. 

Substituting~(\ref{eq:FRHO})-(\ref{eq:F}) in~(\ref{eq:BKG})-(\ref{eq:LHP}), we obtain two systems of integro-differential equations (IDEs) that differ only for the choice of $t_{n_i}$. Note that this substitution introduces a dependence on $\beta$ and $t_n$ because of~(\ref{eq:Gamma}) and the fact that $F(t,t_{n_i})$ depends on $\Gamma$. Note also that the dependence on $t_n$ is equivalent to a dependence on $\alpha$ because the time/temperature relation $Hdt = -dT/T$ implies 
\be \exp\left(4 \int_{t_{\rm eq}}^{t_n} dt' H(t')\right)  = \left(\frac{T_{\rm eq}}{T_n}\right)^4 = \alpha,\ee 
where we used~(\ref{eq:alpha}). 

Now we define 
\begin{equation}
	\label{eq:delta}
	\delta(t,t_{n_i}) \equiv \frac{\rho_R^l(t,t_{n_i})-\rho_R^{b}(t,t_c)}{\rho_R^b(t,t_c)},
\end{equation}
where $\rho^i_R(t,t_{n_i})$, with $i=b,l$,  is the solution of the IDEs for $\rho^i_R$ at time $t$ taking $t_{n_i}$ as the instant when the first bubble forms. 
We recall that in our approach we approximate the energy densities with their
  space averages in separate patches. For example, $\rho_R^l$ is the average of the radiation energy density over the LP under study and  $\rho^b_R$ is the total average of the radiation energy density (over all patches). Therefore, inside a given patch the energy density is approximated with a homogeneous quantity. In this approximation $\delta$ 
   is a measure of the mass excess in the LP.
  Defining $t_{\rm max}$ as the value of  $t$ when $\delta(t,t_{n_i})$ is maximal, we have, given our previous definition of $t_{n_i}^{ \rm PBH}$,
\begin{equation}
	\label{eq:deltaconst}
	\delta(t_{\rm max},t_{n_i}^{\rm PBH})=\delta_c.
\end{equation}
  One can calculate $t_{n_{i}}^{\rm PBH}$ and $t_{\rm max}$ by solving the two systems of IDEs. Our integration strategy is illustrated in Appendix~\ref{appendix}. 
  
  It is important to note that, once a value of $t$ such that $\delta(t,t_{n_i})=\delta_c$ is reached and, therefore, the collapse starts, our model has to be extended to include further inhomogeneities if one wants to follow the collapse process because in the model the patches are separately approximated by homogeneous 
  regions. For $t_{n_i}=t_{n_i}^{\rm PBH}$ such a value of $t$ is $t_{\rm max}$. In particular note that the curvature perturbations cannot be captured here as they represent inhomogeneities over the homogeneous background. Nevertheless, in the definition we use, $\delta$ 
   is a measure of the mass excess in the overdense patch and so it is an estimate of
the $\delta$, which includes non-linear effects, defined in~\cite{Polnarev_2007} (and also in~\cite{Musco_2019}), 
  because we interpret our energy densities as the result of 
  space averages in separate patches. 
   Note that, unlike in other works, the definition of $\delta$ in~\cite{Polnarev_2007} is manifestly gauge invariant and Ref.~\cite{Polnarev_2007} uses a method that provides a non-perturbative description of large amplitude metric perturbations. What we could do in our approximated approach (where separate regions are approximated as homogeneous and isotropic) is to compute if and when $\delta$ exceeds the critical value for gravitational collapse. This means that after $t_{\rm max}$ one must start including inhomogeneities and, therefore, extend the current approach if one wants to study the details of the collapse process. The time $t_{\rm max}$ is really the maximal value of $t$ for which one can trust the present approximation. 

\begin{figure}[t]
\begin{center}
  \includegraphics[scale=0.38]{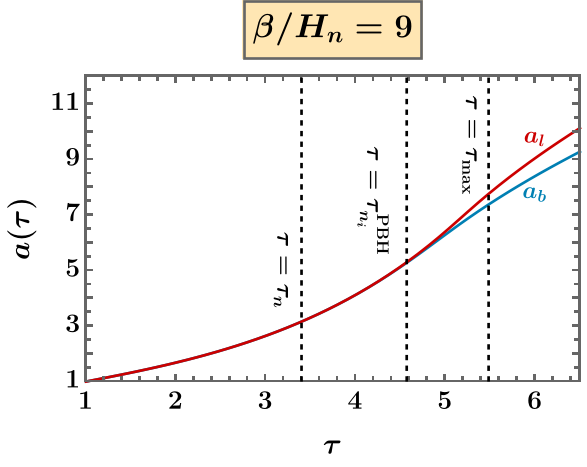}  \hspace{1cm} \includegraphics[scale=0.38]{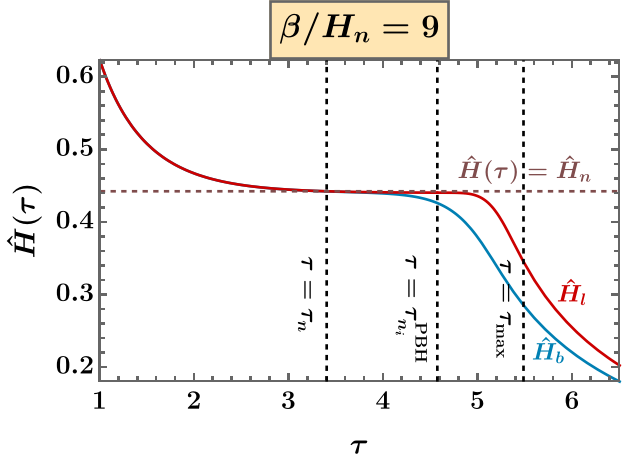}  \\ \vspace{-0.5cm}
  \includegraphics[scale=0.38]{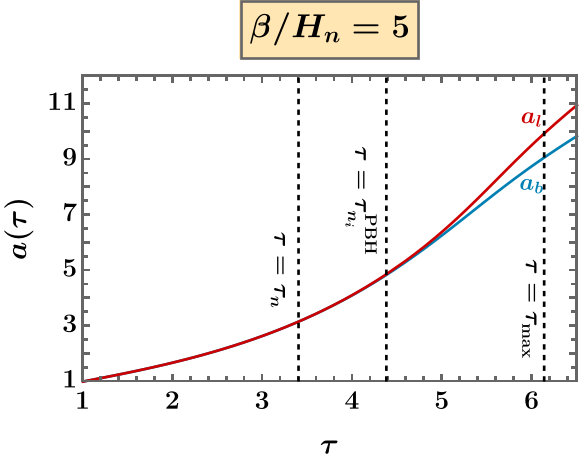}  \hspace{1cm} \includegraphics[scale=0.38]{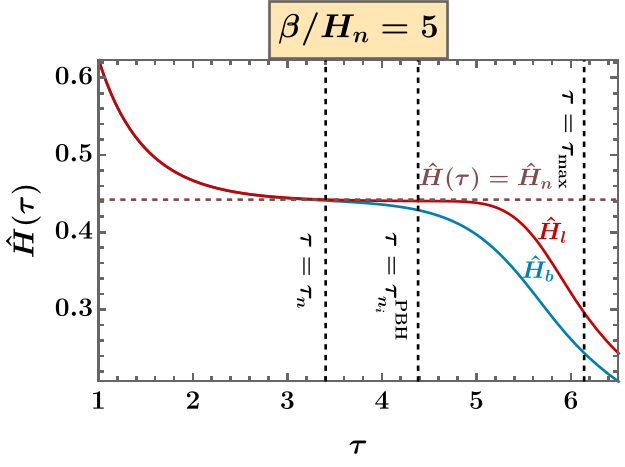}
    \caption{\em  Evolution of the scale factors (left) and Hubble rates (right) for the background and the LP.}  \label{fig:H9plot}
  \end{center}
\end{figure}

\begin{figure}[t]
\begin{center}
\vspace{0.5cm}
  \includegraphics[scale=0.38]{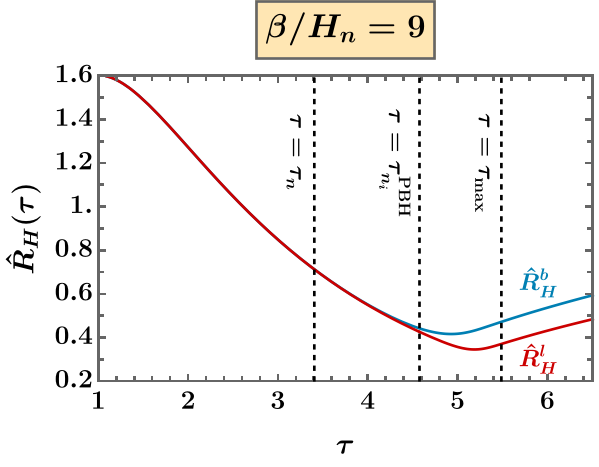}  \hspace{1cm}\includegraphics[scale=0.38]{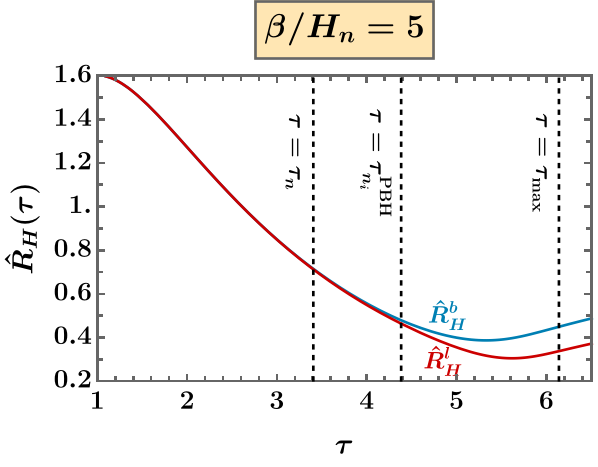} 
    \caption{\em Evolution of the Hubble radius for the background and  the LP.}  \label{fig:RH9plot}
  \end{center}
\end{figure}

\begin{figure}[t]
\begin{center}
  \includegraphics[scale=0.38]{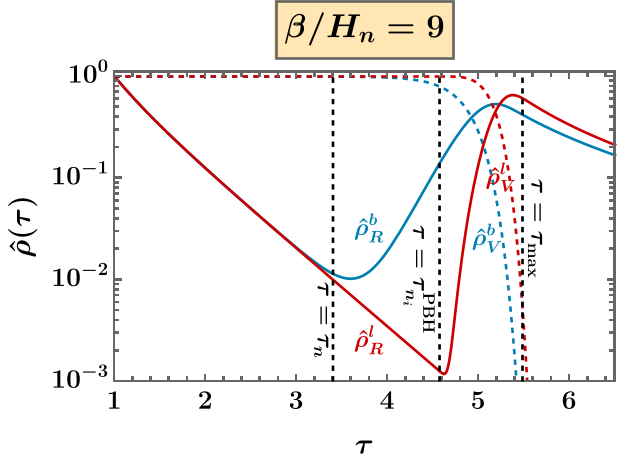}  \hspace{1cm} \includegraphics[scale=0.38]{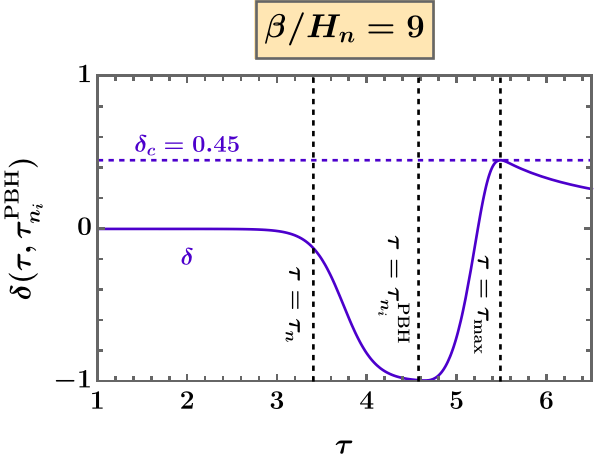} \\
 \vspace{-0.5cm}
   \includegraphics[scale=0.38]{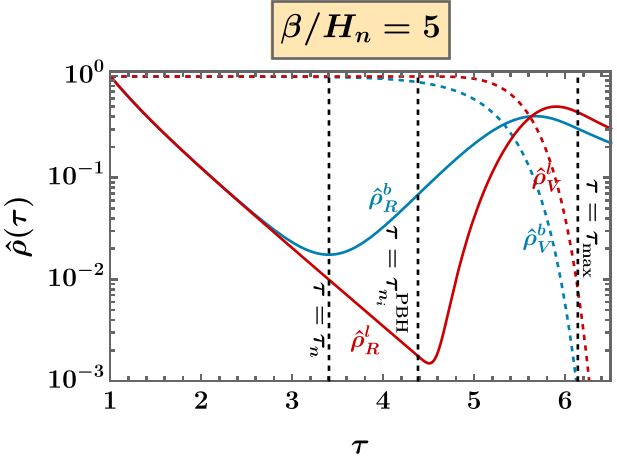}  \hspace{1cm} \includegraphics[scale=0.38]{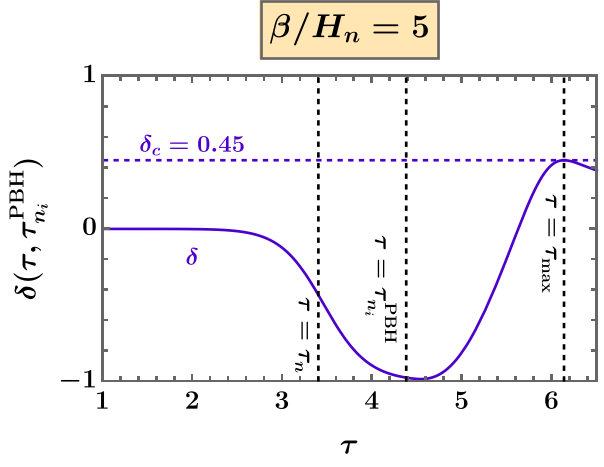} 
    \caption{\em  {\bf Left:} Evolution of the vacuum and radiation energy densities ($\hat\rho_V \equiv \rho_V/\Delta V, \hat\rho_R \equiv \rho_R/\Delta V$) for the background and  the LP.           
		 {\bf Right:} Evolution of the mass excess in the LP,
		  $\delta$. }  \label{fig:Delta9plot}
  \end{center}
\end{figure}

In Figs.~\ref{fig:H9plot},~\ref{fig:RH9plot} and \ref{fig:Delta9plot} we show the time evolution of the scale factors, the Hubble rates, the Hubble radii  and the energy densities   both for the background and the LP, as well as the mass excess in the LP, as functions of time, setting, as examples, the following values of $\beta/H_n$ and $t_n$.
\begin{itemize}
\item $\beta/H_n=9$ and $t_n = 3.4 \,t_{\rm eq}$ (which corresponds to $\alpha\approx 99.7$). For these values we obtained 
\begin{equation}
	t_{n_i}^{\rm PBH} \approx 4.57 \,t_{\rm eq}, \hspace{1cm} t_{\rm max} \approx 5.48 \,t_{\rm eq}.
\end{equation}
\item $\beta/H_n=5$ and $t_n = 3.4 \,t_{\rm eq}$ (which again corresponds to $\alpha\approx 99.7$). For these values we obtained 
\begin{equation}
	t_{n_i}^{\rm PBH} \approx 4.38 \,t_{\rm eq}, \hspace{1cm} t_{\rm max} \approx 6.14 \,t_{\rm eq}.
	\end{equation}
	\end{itemize}
In the plots we used the dimensionless variables \be \hat H \equiv H t_{\rm eq},  
\qquad \hat R_{H}\equiv \frac1{a \hat{H}}, \qquad 
\hat\rho \equiv \frac{\rho}{\Delta V}, \qquad 
\tau\equiv \frac{t}{t_{\rm eq}}. \label{NoDimT}\ee
In the left plots of Fig.~\ref{fig:H9plot}  we can see that the scale factors evolve in the same way at the beginning, but then inflation ends first in the background and only later in the LP. The right plots of Fig.~\ref{fig:H9plot} show the evolution of the two corresponding Hubble rates and we see that, at the beginning they are equal, up to the vacuum domination era, when an inflationary expansion takes place and the Hubble rate  equals a constant, $H_I$; in particular we see $H_I$ is very close to $H_n$ in that period. Then the inflationary expansion stops and  the Hubble rates start to decrease first in the background and only later in the LP.
The two different durations of the inflationary expansion for the background and the LP are also clear in Fig.~\ref{fig:RH9plot}, showing the corresponding Hubble radii. 
In the left plots of Fig.~\ref{fig:Delta9plot} we see that the radiation energy density decreases initially both in the background and in the LP, then after $\tau_n\equiv t_n/t_{\rm eq}$, the quantity $\hat{\rho}^{b}_{R}\equiv \rho^{b}_{R}/\Delta V$ starts to increase because of the bubble nucleation and then starts to decrease as expected for a radiation energy density. The same is true for $\hat{\rho}^l_R\equiv \rho^{l}_{R}/\Delta V$, which, however, starts to increase only after $\tau_{n_i}^{\rm PBH}\equiv t_{n_i}^{\rm PBH}/t_{\rm eq}$.
In the right plots of Fig.~\ref{fig:Delta9plot} we can see the behavior of $\delta$, which reaches its peak, $\delta_c$, at $\tau_{\rm max}\equiv t_{\rm max}/t_{\rm eq}$.

Figs.~\ref{fig:H9plot},~\ref{fig:RH9plot} and \ref{fig:Delta9plot} also show the dependence on $\beta/H_n$. Decreasing $\beta/H_n$ the time scale  of the evolution becomes larger (as expected) and $\tau_{n_i}^{\rm PBH}$ becomes smaller. In other words, for smaller values of $\beta/H_n$ the minimal $\tau_{n_i}$ required to reach a given mass excess in the LP
($\delta_c=0.45$ in the plots) becomes smaller. So the PBH production becomes more efficient decreasing $\beta/H_n$.

Now, note that the exponential in~(\ref{eq:F}) represents the probability that a given point in space remains in the false vacuum at time $t$, {\it assuming} that no bubble formed before $t_{n_i}$. One then needs to know the probability $\mathcal{P}_{\rm surv}(t_{n_i},t_{\rm max})$ that in a LP at time $t_{\rm max}$ (the maximal time before the collapse for $t_{n_i}=t_{n_{i}}^{\rm PBH}$) no bubble formed before $t_{n_i}$. This probability was originally determined for a Hubble patch of radius $H^{-1}(t_{\rm max})$ (the Hubble radius at $t_{\rm max}$) in~\cite{Kodama:1982sf} and  re-derived in~\cite{Gouttenoire:2023naa}:
\begin{equation}
	\label{eq:Psurv}
	\mathcal{P}_{\rm surv}(t_{n_i},t_{\rm max}) = \text{exp}\left[-\int^{t_{n_i}}_{t_c}dt'\Gamma (t')a(t')^3 V(t',t_{\rm max})\right],
\end{equation}
where
\begin{equation}
	V(t',t_{\rm max}) = \frac{4\pi}{3}(R_{H}(t_{\rm max})+r(t_{\rm max},t'))^3, \qquad R_H(t)\equiv\frac1{a(t)H(t)}
\end{equation} 
and all quantities refer to the LP.

The exponential in~(\ref{eq:Psurv}) decreases very rapidly increasing $t_{n_i}$, so the collapse probability is estimated as
\begin{equation}
	\label{eq:Pcoll}
	\mathcal{P}_{\rm coll}\equiv \mathcal{P}_{\rm surv}(t_{n_i}^{\rm PBH},t_{\rm max}).
\end{equation}
The exponent in~(\ref{eq:Psurv}) decreases very rapidly also increasing $\beta$, as clear from~(\ref{eq:Gamma}) (justified in Subsec.~\ref{Time dependence of the false-vacuum decay rate} for RSB). So, it is again clear that smaller values of $\beta$ corresponds to a larger PBH production. 
 
  We find that our numerical results (see Appendix~\ref{appendix} for our method of integration) are well described  for $\alpha \gtrsim10^2$ by the following fitting function:
\begin{equation}
	\label{eq:collfit}
	\mathcal{P}_{\rm coll} \approx \text{exp}\left[-a_{\mathcal{P}}\left(\frac{\beta}{H_n}\right)^{b_\mathcal{P}} (1+\delta_c)^{c_{\mathcal{P}}\frac{\beta}{H_n}}\right],
\end{equation} 
with $a_{\mathcal{P}}\approx 0.5646, b_{\mathcal{P}}\approx1.266,  c_{\mathcal{P}}\approx 0.6639$ in agreement with the result in~\cite{Gouttenoire:2023naa}, varying $\delta_c$ around $0.45$. 
Note that the large-$\alpha$ limit ($\alpha \gtrsim10^2$) is all we need to apply the late-blooming mechanism to supercooled phase transitions, such as those with RSB~\cite{Salvio:2023qgb,Salvio:2023ext}. 

The rest of the derivation of the PBH fraction of DM, $f_{\rm PBH}$, and the PBH mass, $M_{\rm PBH}$, parallels that  in~\cite{Escriva:2021pmf,Gouttenoire:2023naa}, to obtain
\begin{equation}
	\label{eq:fPBH}
	f_{\rm PBH}\approx  \frac{\mathcal{P}_{\rm coll}}{6.0\times 10^{-12}}\,  \frac{T_{\rm eq}}{500 \text{GeV}}, \qquad M_{\rm PBH}\approx M_{\bigodot} \bigg(\frac{20}{g_*(T_{\rm eq})}\bigg)^{1/2}\bigg(\frac{140\,\text{MeV}}{T_{\rm eq}}\bigg)^2,
\end{equation}
where $M_{\bigodot}$ is the solar mass.

On the other hand, we find that the initial spin of the PBH at the time of creation\footnote{The authors of Ref.~\cite{Banerjee:2023qya} have studied the initial spin of PBHs created from FOPTs. However, the curvature perturbation considered in that article does not consider very high values of $\alpha$ and hence is not compatible with the RSB scenario.} can be expressed as 
\begin{align}
\sqrt{\langle a_*^2\rangle} \approx \dfrac{2.1\times 10^{-3}}{23.484-1.25\log_{10}(f_{\mathrm{PBH}})-1.25\log_{10}\left(\dfrac{\Omega_{\mathrm{CDM}}}{0.26}\right)-0.625\log_{10}\left(\dfrac{M_{\rm PBH}}{10^{15}\mathrm{~g}}\right)}.
\label{astarfinal}
\end{align}
Here $\sqrt{\langle a_*^2\rangle}$ is the root mean square (RMS) value of the dimensionless Kerr parameter $a_*$ and $\Omega_{\mathrm{CDM}} \approx 0.26$.   In the Boyer-Lindquist coordinates the Kerr parameter of a rotating black hole is given in terms of its mass and absolute value $J$ of angular momentum  by (for our choice of units)  $a_K=J/M_{\rm PBH}$, which has the dimension of the inverse of a mass. The dimensionless Kerr parameter $a_*$ can be expressed as\footnote{In the international system of units $a_K=J/(M_{\rm PBH}c)$ and $a_*=a_Kc^2/(G_NM_{\rm PBH})$, where $c$ is the speed of light.} $a_*=a_K/(G_NM_{\rm PBH})$, where $G_N$ is Newton's constant.  We will refer to $a_*$ as the spin of the black hole. 

The expression in~(\ref{astarfinal}) has been derived by using the result of Ref.~\cite{Banerjee:2024nkv}  for $\sqrt{\langle a_*^2\rangle}$ and the fact that the PBHs are produced in the radiation-dominated era after the RSB PT with a mass given by the mass inside the sound horizon $H^{-1}/\sqrt{3}$~\cite{Gouttenoire:2023naa}. 
Ref.~\cite{Banerjee:2024nkv} followed in turn the treatment of Refs.~\cite{DeLuca:2019buf,Harada:2020pzb}, except that~\cite{Banerjee:2024nkv} used a  power spectrum of the curvature
perturbation that is more suitable for first-order phase transitions. 
However, as mentioned in Ref.~\cite{DeLuca:2019buf}, the generation of the PBH spin is only generated if two conditions are satisfied. One is that the overdense regions, which eventually collapse, must be non-spherical. The second condition is that the velocity shear must be misaligned with the collapsing object's inertia tensor. 

The collapsing regions in the mechanism we focus on are nearly-spherical, but not entirely spherical, which satisfy the first condition. This is due to the fact that there exists many true vacuum bubbles in a delayed patch when the collapse takes place in the radiation-dominated era.

The second condition is satisfied as a mean statistical property~\cite{Bardeen:1985tr,HeavensPeacock}, where Gaussian distributions of the relevant quantities are assumed. The plausibility of this assumption comes from the fact that the PBH abundance from the present mechanism is the result of the contribution of a large number of patches. 
Calculating the angular momentum of each collapsing object would be more involved and is out of the scope of the present work.

\section{General supercooled phase transitions from RSB and PBHs}\label{General supercooled phase transitions from RSB and PBHs}

We now apply the PBH production mechanism of Sec.~\ref{A fresh look at the late-blooming mechanism} to 
the class of theories (or, more generally, sectors) where  symmetries are broken and masses are generated radiatively~\cite{Coleman:1973jx,Gildener:1976ih}, at least  in some regions of their parameter space. 
As mentioned in the introduction, these RSB systems feature supercooled FOPTs and can be analysed through perturbative methods. In the first part of this section we recap the features of RSB and the corresponding PTs that are needed to understand the original results of this section (Subsecs~\ref{Time dependence of the false-vacuum decay rate} and \ref{PBH abundance, mass and initial PBH spin}). For the proof of any non-trivial statement in this first part  we refer to Refs.~\cite{Salvio:2023qgb,Salvio:2023ext}.

RSB can occur when the classical theory features a flat direction in the space of scalar fields\footnote{We consider a generic number $N_S$ of scalars ($a=1, ... ,N_S$), so we can take the $\phi_a$ real without loss of generality.} $\phi_a$. We can parameterize such direction as $\phi_a = \nu_a \chi$, where $ \nu_a$ are the components of a unit vector $\nu$, i.e.~$ \nu_a  \nu_a =1$, and $\chi$ is a single scalar field. Inserting this parameterization in the generic no-scale potential\footnote{Unlike the matter sector under study, gravity is assumed to be the Einsteinian one at the energies that are relevant for this work. It is possible, however, to construct a classically scale-invariant theory of all interactions, gravity included, where scale invariance is broken by dimensional transmutation~\cite{Salvio:2014soa,Kannike:2015apa,Salvio:2017qkx,Salvio:2017xul,Salvio:2019wcp,Alvarez-Luna:2022hka} at energies  above those of interest here.}.
\be  \frac{\lambda_{abcd}}{4!} \phi_a\phi_b\phi_c\phi_d, \label{Vns}\ee
where $\lambda_{abcd}$ are the quartic couplings, one obtains the RG-improved potential of $\chi$:
\be V(\chi) = \frac{\lambda_\chi (\mu)}{4}\chi^4, \qquad (\lambda_\chi(\mu) \equiv\frac1{3!} \lambda_{abcd}(\mu) \nu_a \nu_b \nu_c \nu_d), \label{Vvarphi}\ee 
where $\mu$ is the renormalization scale.
Including the one-loop correction, the quantum effective potential reads
 \be V_q(\chi) = \frac{\bar \beta}4\left(\log\frac{\chi}{\chi_0}-\frac14\right)\chi^4,\label{CWpot}\ee
 where $\chi_0$ is a free non-vanishing parameter with the dimension of energy,
 \be \bar\beta \equiv \left[\mu\frac{d\lambda_{\chi}}{d\mu}\right]_{\mu=\tilde\mu} \label{betabardef}\ee
and $\tilde\mu$ is the energy scale where the flat direction appears, $\lambda_\chi(\tilde\mu)=0$. When $\bar\beta>0$ the potential in~(\ref{CWpot}) has a minimum at $\chi_0\neq 0$ and this can trigger RSB. Including thermal one-loop effects, one always obtains an effective potential that admits the following large-supercooling expansion (barring an additive field-independent constant)
\be \bar V_{\rm eff}(\chi,T) = \frac{m^2(T)}{2} \chi^2-\frac{k(T)}{3}\chi^3-\frac{\lambda(T)}{4} \chi^4 + ... , \label{barVnlo}\ee
where
\be m^2(T) \equiv \frac{g^2 T^2}{12},\qquad k(T)\equiv \frac{\tilde g^3 T}{4\pi}, \qquad \lambda(T) \equiv \bar\beta \log\frac{\chi_0}{T}, \label{mklambdaDef}\ee
 \be g^2 \equiv \sum_b n_bm_b^2(\chi)/\chi^2+\sum_f m^2_f(\chi)/\chi^2, \qquad \tilde g^3 \equiv \sum_b n_bm_b^3(\chi)/\chi^3. 
\label{M2g2gt3def}\ee
Here the $m_b$ and $m_f$ are the background-dependent bosonic and fermionic masses, respectively, the sum over $b$ runs over all bosonic degrees of freedom and $n_b=1, 3$ for a scalar and a vector degree of freedom, respectively.  The quantities $g$ and $\tilde g$ turn out to be $\chi$ independent and they always satisfy $\tilde g\leq g$. We don't restrict the number of bosons and fermions and we work with Weyl spinors for generality. This allows us to be model independent.

When the parameter
\be \epsilon\equiv  \frac{g^4}{6\bar\beta \log\frac{\chi_0}{T}}
 \label{CondConv}\ee
 is small for the relevant temperatures, which occurs for a large-enough hierarchy $T\ll \chi_0$ (large-enough supercooling), for our purposes one can treat the cubic-in-$\chi$ term in~(\ref{barVnlo}) perturbatively, together with the additional terms in the dots. This perturbative approach is named the {\it``supercool expansion"}. For example, setting those terms to zero one obtains the supercool expansion at leading order (LO), where the full effective potential can be approximated by
 \be \bar V_{\rm eff}(\chi,T) \approx \frac{m^2(T)}{2} \chi^2-\frac{\lambda(T)}{4} \chi^4. \label{Vlo}\ee
 When there are several degrees of freedom, say $N$, with dominant couplings (all of the same order of magnitude) to $\chi$ the supercool expansion is actually valid even for larger values of $\epsilon$. This is because $\tilde g^3/g^3\lesssim1/\sqrt{N}$.  The  supercool expansion becomes even more accurate if there are much more fermions  than bosons with dominant couplings  to $\chi$ as only bosons contribute to $\tilde g$. 
 
 When the supercool expansion breaks down one can still perform a perturbative treatment of this general system as long as $\epsilon$ is not large, say at most {\it of order} 1, i.e.~$\epsilon\lesssim 1$, which still requires some supercooling. In this case the cubic-in-$\chi$ term in~(\ref{barVnlo}) should be treated exactly, like the quadratic and quartic ones. 
 Instead, the terms in the dots of~(\ref{barVnlo}) can still be treated perturbatively. This perturbative approach is named the {\it``improved supercool expansion"} and has a larger regime of validity of the supercool expansion previously defined. In the improved supercool expansion one can approximate
 \be \bar V_{\rm eff}(\chi,T) \approx \frac{m^2(T)}{2} \chi^2-\frac{k(T)}{3}\chi^3-\frac{\lambda(T)}{4} \chi^4, \label{improLO}\ee 
treating all coefficients $m$, $k$ and $\lambda$ exactly.

In both the supercool expansion and the improved version the decay rate of the false vacuum is dominated by the time-independent bounce. The corresponding three-dimensional action $S_3$ is given by
\be S_3\approx c_3 \frac{m}{\lambda}, \qquad c_3=18.8973...\label{S3c3}\ee 
in the supercool expansion at LO, while, in the improved approximation of~(\ref{improLO}), $S_3$   is well fitted by
\be S_3 \approx 27\pi m^3 \frac{1+\exp(-k/(m\sqrt{\lambda}))}{2k^2+9\lambda m^2} \label{S3k} \ee
for the values of the parameters that are relevant for us\footnote{The derivative corrections to the canonical kinetic term of $\chi$ and the higher-derivative corrections turn out to be negligible in this  approximation.}. 
The simple form in~(\ref{S3c3}) allows for an analytical determination of  $T_n$
\be T_n\approx \chi_0\exp\left(\frac{\sqrt{c'^2-16a'}-c'}8\right), \label{appTn}\ee
where
\be a'\equiv \frac{c_3g}{\sqrt{12}\bar\beta}, \quad c'\equiv 4\log\frac{4\sqrt{3}\bp}{\sqrt{\bar\beta}\,\chi_0} \label{caDef}\ee
and of $\beta/H_n$
\be \frac{\beta}{H_n} \approx \frac{a'}{\log^2(\chi_0/T_n)} -4. \label{betaH2} \ee
 Numerical calculations are instead used to obtain $T_n$ in the improved approximation of~(\ref{improLO}). In the present paper we performed a refined numerical calculation for better accuracy.

The great advantage of these expansions is that they allows us to have a model-independent approach: everything can be expressed in terms of few parameters ($\chi_0$, $\bar\beta$, $g$, $\tilde g$, ...) that are computable once the model is specified. We will then use such a model-independent approach in this paper.

\subsection{Time dependence of the false-vacuum decay rate}\label{Time dependence of the false-vacuum decay rate}

Since the time-independent bounce always dominates for RSB PTs in both the supercool and improved supercool expansions,  $\Gamma$ can be approximated by
\be  \Gamma\approx T^4\exp(-S_3/T). \label{Gamma3}\ee
Also, in theories with RSB, using~(\ref{CWpot}),  
 \be \Delta V = \frac{\bar\beta \chi_0^4}{16}, \ee
 which leads to
 \be H_I = \frac{\sqrt{\bar\beta} \chi_0^2}{4\sqrt{3}\bp}, \qquad T_{\rm eq}^4 = \frac{15\bar{\beta}\chi_{0}^4}{8\pi^2 g_{*}(T_{\text{eq}})}.\label{HITeq}\ee

We now prove that the approximation for the time dependence of $\Gamma$ in~(\ref{eq:Gamma}), which is widely used in the context of PBH production from FOPTs, is valid and it proves to be increasingly accurate as the degree of supercooling becomes larger.

To understand this result let us consider the phase of the evolution of the universe since a time close to $t_n$. We can do so because vacuum decay is not effective at much earlier times. So, we can approximate $H(t)\approx H_I$ (constant in time) since the spacetime exponentially expand at those times before vacuum decay occurs. Note that here we have used the presence of supercooling, which tells us that the energy density is dominated by the vacuum contribution at $t_n$. 
Using $dt = -dT/(TH) \approx -dT/(TH_I)$ we find
\be T(t) \approx T_n e^{-H_I(t-t_n)}. \label{tTcon} \ee
Let us consider now the supercool expansion at LO; as we will explain, the extension to higher orders and even to the improved supercool expansion can be easily obtained. In this approximation
\be \frac{S_3}{T}\approx c_3 \frac{m}{T\lambda} = \frac{c_3 g}{\sqrt{12}\bar\beta \log\frac{\chi_0}{T}} \equiv \frac{a'}{X+\log\frac{T_n}{T}}\approx \frac{a'}{X+H_I(t-t_n)},\ee
where we have defined $X\equiv \log\frac{\chi_0}{T_n}$ and in the last step we used~(\ref{tTcon}).
  Inserting now this result and~(\ref{tTcon}) in~(\ref{Gamma3}) gives
\be  \Gamma(t) \approx T_n^4\exp\left(-\frac{a'}{X+H_I(t-t_n)}-4 H_I(t-t_n)\right). \ee
Expanding for $t$ around $t_n$ the first term in the  exponent, 
\be \frac{1}{1+\frac{H_I(t-t_n)}{X}} =  1 -\frac{H_I(t-t_n)}{X}+\left(\frac{H_I(t-t_n)}{X}\right)^2+... + (-1)^k\left(\frac{H_I(t-t_n)}{X}\right)^k + ...  \label{TayE}\ee 
we can clearly see that the constant and ${\cal O}(t-t_n)$ terms reproduce the right-hand side of~(\ref{eq:Gamma}) because $a'/X$ equals $S_3/T$ at $t=t_n$ and 
\be \frac{a' H_I}{X^2} - 4H_I \approx \beta \ee
according to~(\ref{betaH2}). The order of magnitude of the generic ${\cal O}(t-t_n)^k$ term in~(\ref{TayE}) can be estimated by noting that
\be \frac{H_I(t-t_n)}{X} = \frac{\gamma(\tau-\tau_n)}{\sqrt{3}X}, \ee
where we used~(\ref{gamma}),~(\ref{teqHeq}) and~(\ref{HeqDef}) and so $H_I =H_{\rm eq}/\sqrt{2} = \gamma/(\sqrt{3} t_{\rm eq})$. Now, in the supercool expansion $X$ is rather large. Indeed, $X$ is typically more than an order of magnitude larger than $1/\epsilon$ evaluated at $T_n$ (as $\bar\beta$ is typically of order $g^4/(4\pi)^2$) and $\sqrt{3} X/\gamma$ is even larger (see~(\ref{gamma})). 
So, as long as one is interested in time differences $\tau-\tau_n$ that are small compared to $\sqrt{3} X/\gamma$, as we do\footnote{Indeed, in the example of Figs.~\ref{fig:H9plot},~\ref{fig:RH9plot} and \ref{fig:Delta9plot} the relevant values of $\tau-\tau_n$ are smaller than $\approx 3$. 
},~(\ref{eq:Gamma}) is a good approximation, which becomes increasingly accurate as the supercooling (namely $X$) increases. 

Note that this is an order-of-magnitude estimate and, as such, it remains valid considering the higher-order corrections in the supercool expansion. Moreover, even in the improved supercool expansion, since $\epsilon$ is at most of order 1 and $\bar\beta$ is typically of order $g^4/(4\pi)^2$, the suppression factor $\gamma/(\sqrt{3}X)$ is rather small, typically at most of order of few \%. We have also numerically checked this with the fit in~(\ref{S3k}), which is valid in the improved supercool expansion where the effective potential is approximated as in~(\ref{improLO}). 

\subsection{PBH abundance, mass and initial spin}\label{PBH abundance, mass and initial PBH spin}

We now use the previous results to study the PBH abundance, mass and initial spin corresponding to a generic theory with RSB.

We compute $f_{\rm PBH}$ by using the first expression in~(\ref{eq:fPBH}) and~(\ref{eq:collfit}) and inserting the value of $\beta/H_n$ previously determined: in the approximation of~(\ref{Vlo}) we use the formula for $\beta/H_n$ given in~(\ref{appTn})-(\ref{betaH2}), while in the approximation of~(\ref{improLO}) we use the numerical approach illustrated in Ref.~\cite{Salvio:2023ext} (but performed here with better accuracy).

In theories with RSB, on the other hand, $M_{\rm PBH}$  can be obtained by combining~(\ref{eq:DV}) with the second expression in~(\ref{eq:fPBH}):
\be M_{\rm PBH}\approx M_{\bigodot} \bigg(\frac{2\pi^2}{3\bar\beta}\bigg)^{1/2}\bigg(\frac{280\,\text{MeV}}{\chi_0}\bigg)^2 \ee
with no dependence on $g_*$. On the other hand, $f_{\rm PBH}$ has a dependence on $g_*$, due to the dependence on $g_*$ of $T_{\rm eq}$. However, this dependence is very weak, being proportional to $1/\sqrt[4]{g_*}$. In the numerical results below we will use a reference value $g_*=10^2$ (around the SM value); but   variations keeping this order of magnitude produce very small modifications.

As far as the initial spin is concerned, $\sqrt{\langle a_*^2\rangle}$ can be obtained by combining the results for $f_{\rm PBH}$ and $M_{\rm PBH}$ discussed above with the expression in~(\ref{astarfinal}).

We now illustrate the dependence of $\beta/H_n$, $f_{\rm PBH}$, $M_{\rm PBH}$ and $\sqrt{\langle a_*^2\rangle}$ on the parameters of the model-independent approach valid for RSB theories with large-enough supercooling.

\begin{figure}[ht!]
		\includegraphics[width=0.41\linewidth]{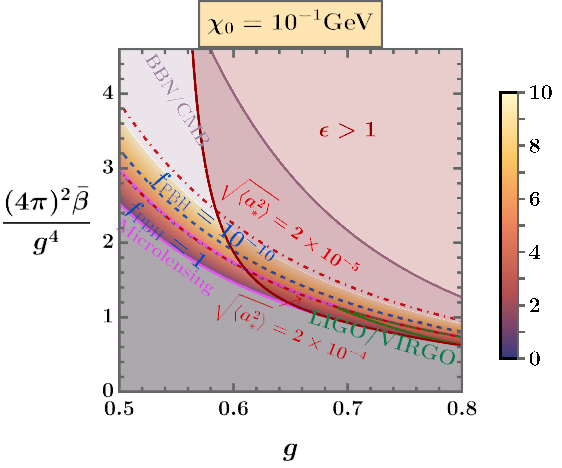} \hspace{0.4cm}
		\includegraphics[width=0.41\linewidth]{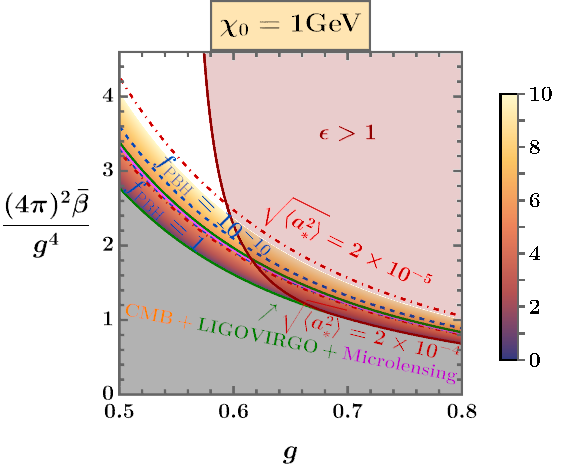} 
	\\
	\includegraphics[width=0.4\linewidth]{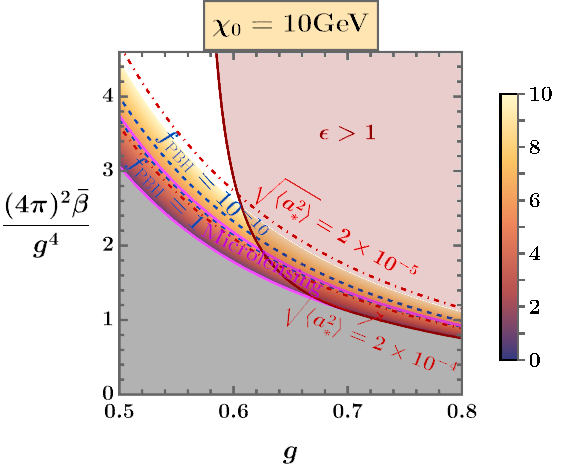} \hspace{0.5cm}
		\includegraphics[width=0.4\linewidth]{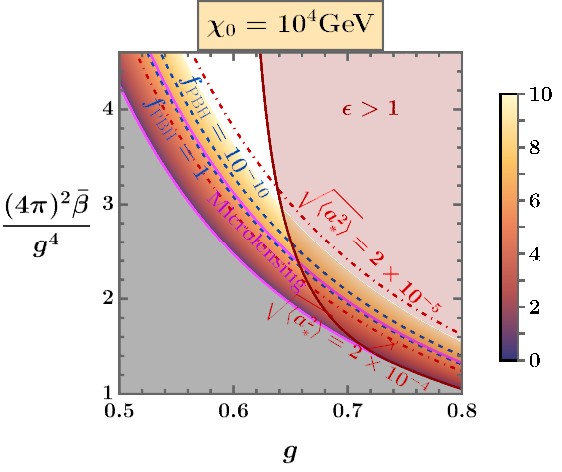}
\\
		\includegraphics[width=0.4\linewidth]{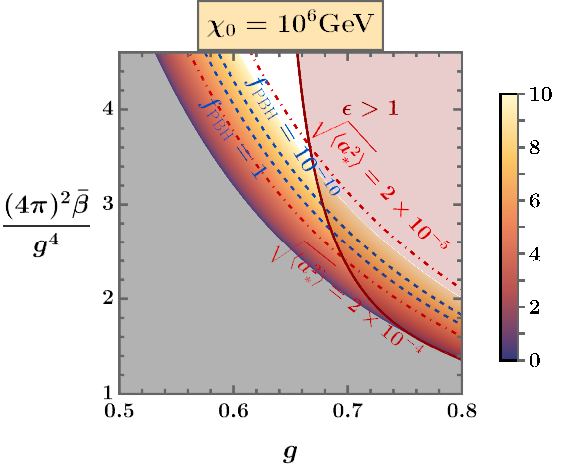} \hspace{0.5cm}
		\includegraphics[width=0.4\linewidth]{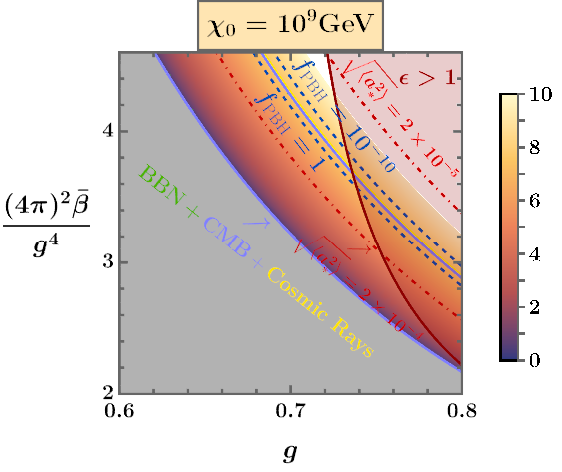}
	\caption{\em  $\beta/H_n$  calculated using   the supercool expansion at LO for fixed values of $\chi_0$. }
		\label{fig:plote<<1}
\end{figure}

%

\begin{figure}[ht!!!!!!]
	\includegraphics[width=0.75\linewidth]{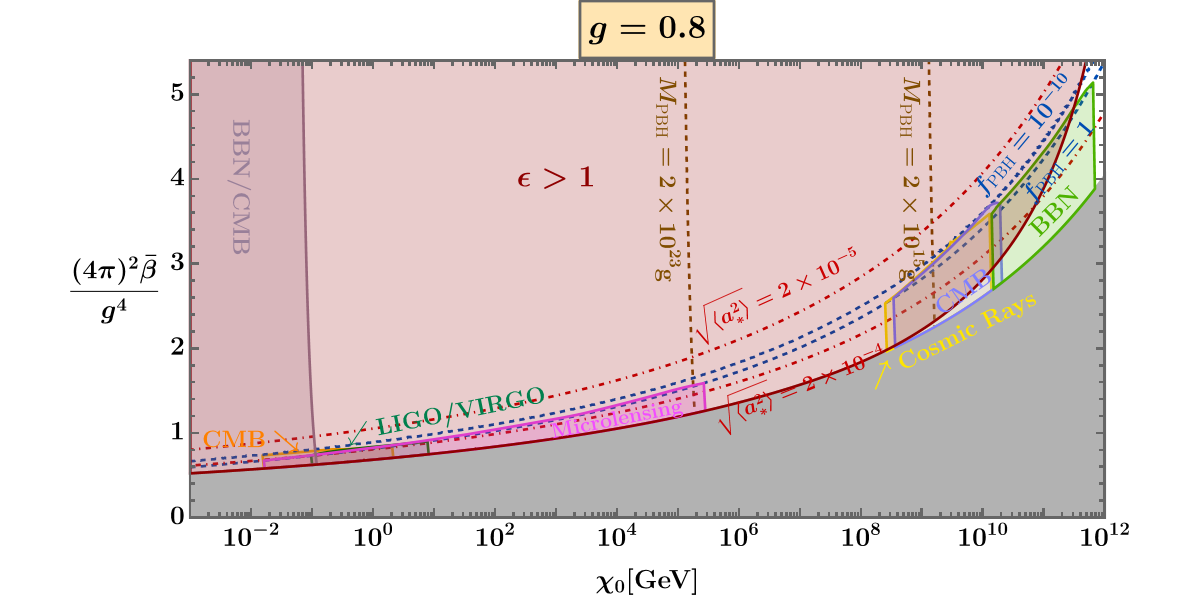}
	
	\vspace{-0.45cm}
	
		\includegraphics[width=0.77\linewidth]{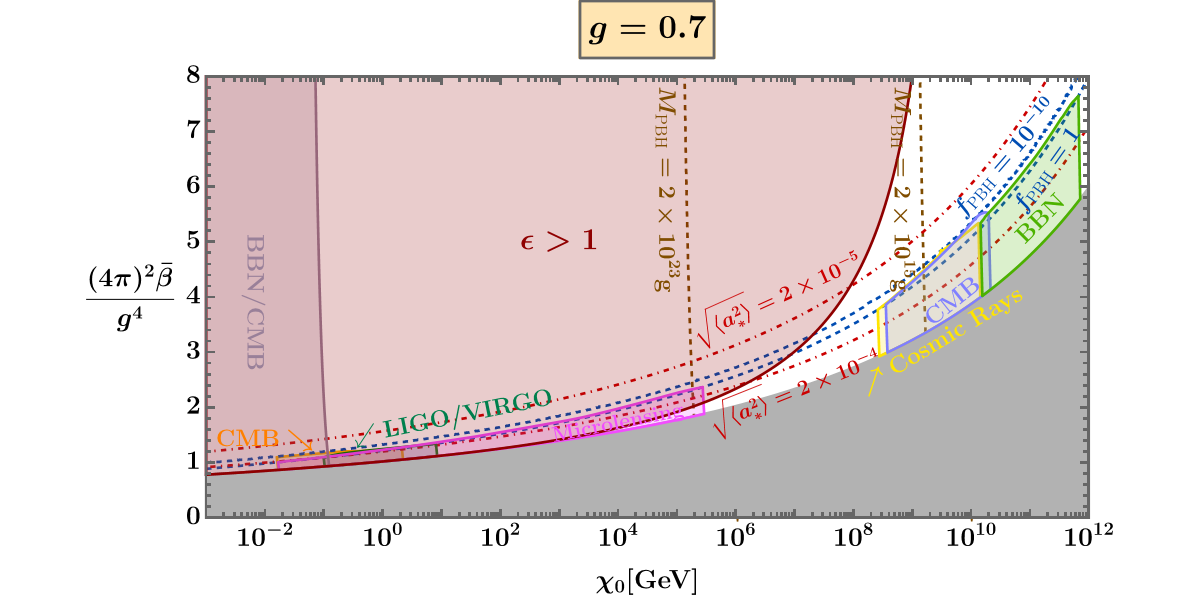} 
		\caption{\em $f_{\rm PBH}$, $M_{\rm PBH}$ and  $\sqrt{\langle a _*^{1/2}\rangle}$
	 calculated with the supercool expansion at LO for fixed values of $g$.}
	\label{fig:ychi0807SE}
\end{figure}

		\begin{figure}[ht!!!!!!]
\vspace{-0.45cm}
		
		\includegraphics[width=0.75\linewidth]{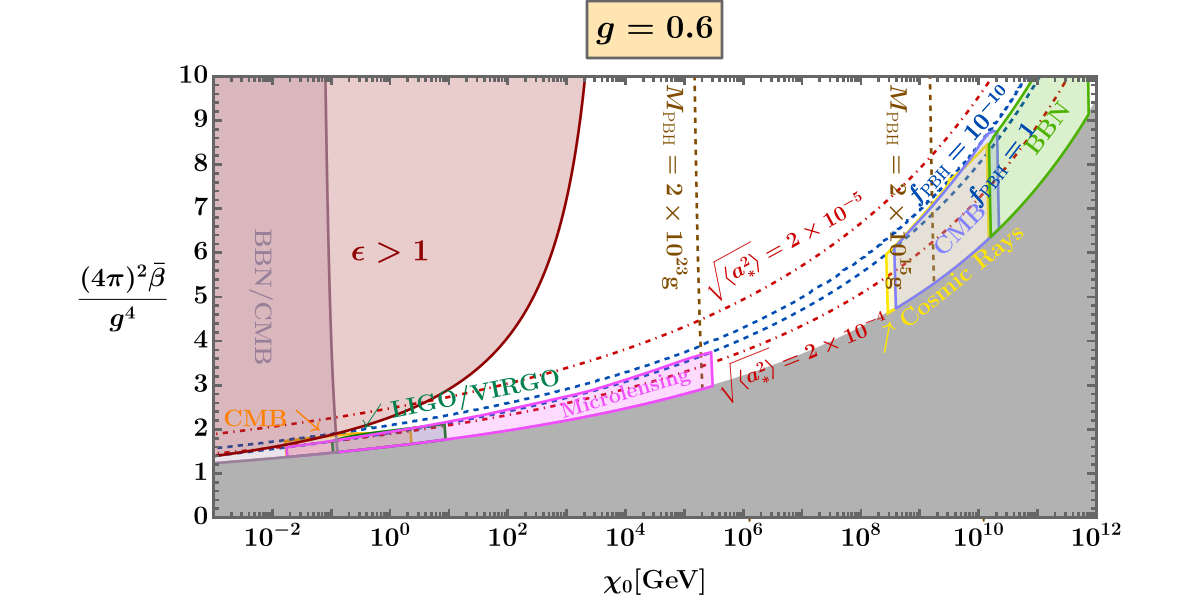}
		
		\vspace{-0.45cm}
		
		\includegraphics[width=0.75\linewidth]{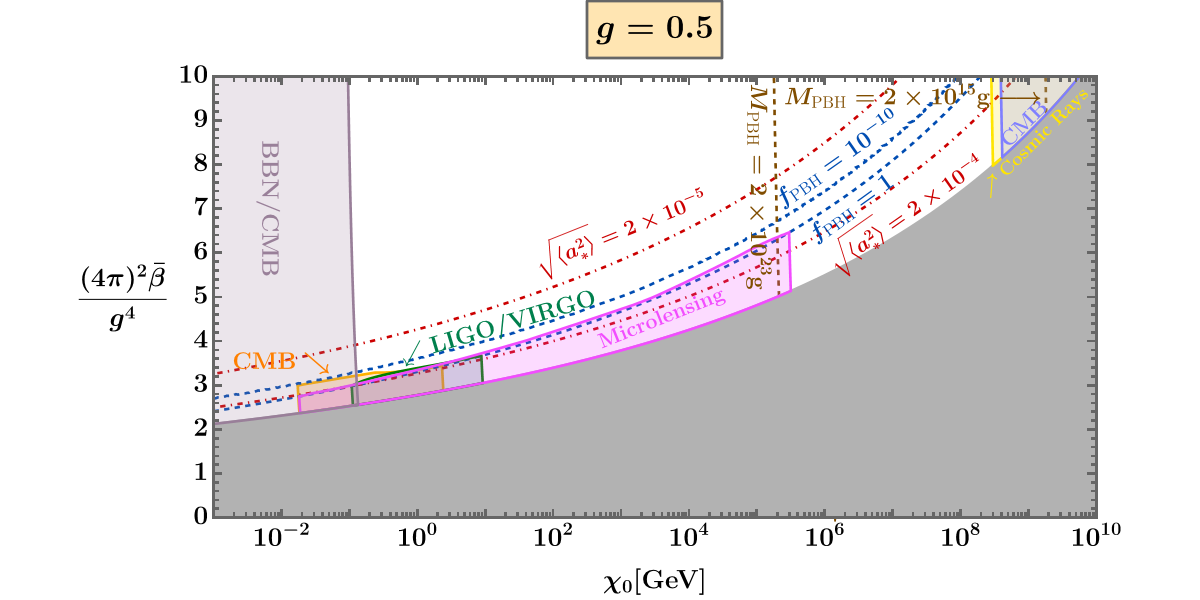}
	\caption{\em Like in Fig.~\ref{fig:ychi0807SE} but for smaller values of $g$.}
	\label{fig:ychi0807SEp}
\end{figure}

\begin{figure}[ht!]
		\includegraphics[width=0.41\linewidth]{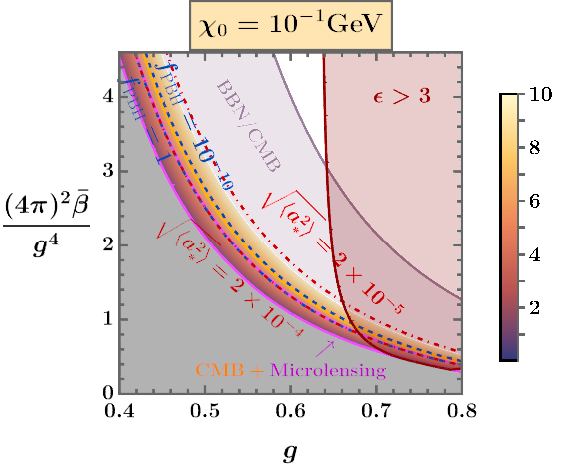} \hspace{0.4cm}
		\includegraphics[width=0.41\linewidth]{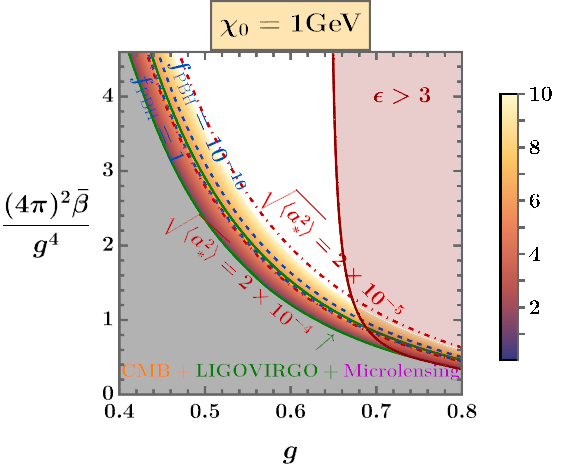} 
	\\
	\includegraphics[width=0.4\linewidth]{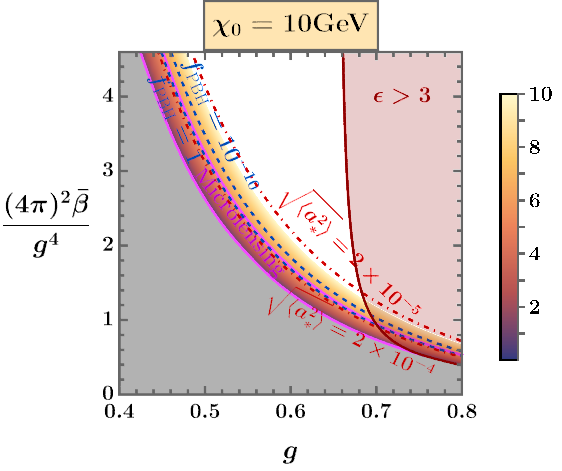} \hspace{0.5cm}
		\includegraphics[width=0.4\linewidth]{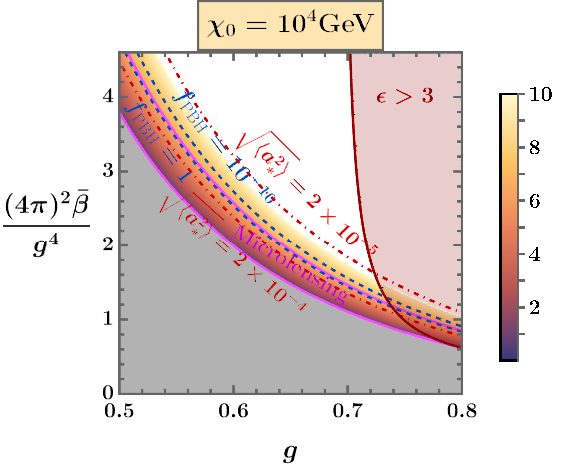}
\\
		\includegraphics[width=0.4\linewidth]{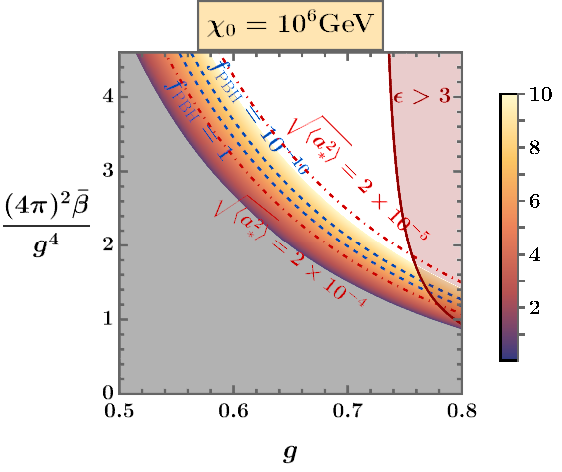} \hspace{0.5cm}
		\includegraphics[width=0.4\linewidth]{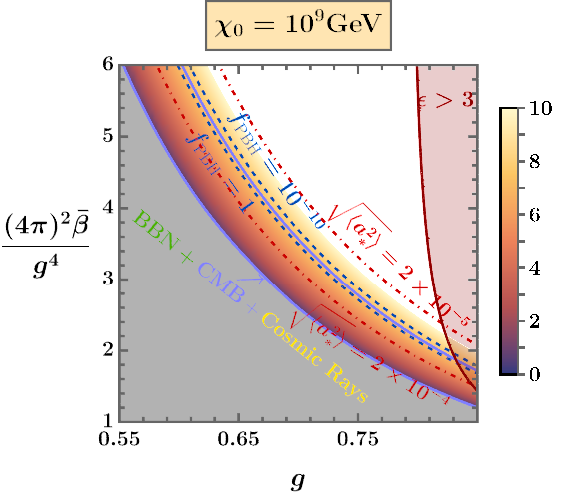}
	\caption{\em  $\beta/H_n$  calculated using the improved supercool expansion where the effective potential is approximated as in~(\ref{improLO}) with $g=\tilde{g}$ and  for  fixed values   of $\chi_0$. }
		\label{fig:fixedchi}
\end{figure}


\begin{figure}[ht!]
	\includegraphics[width=0.75\linewidth]{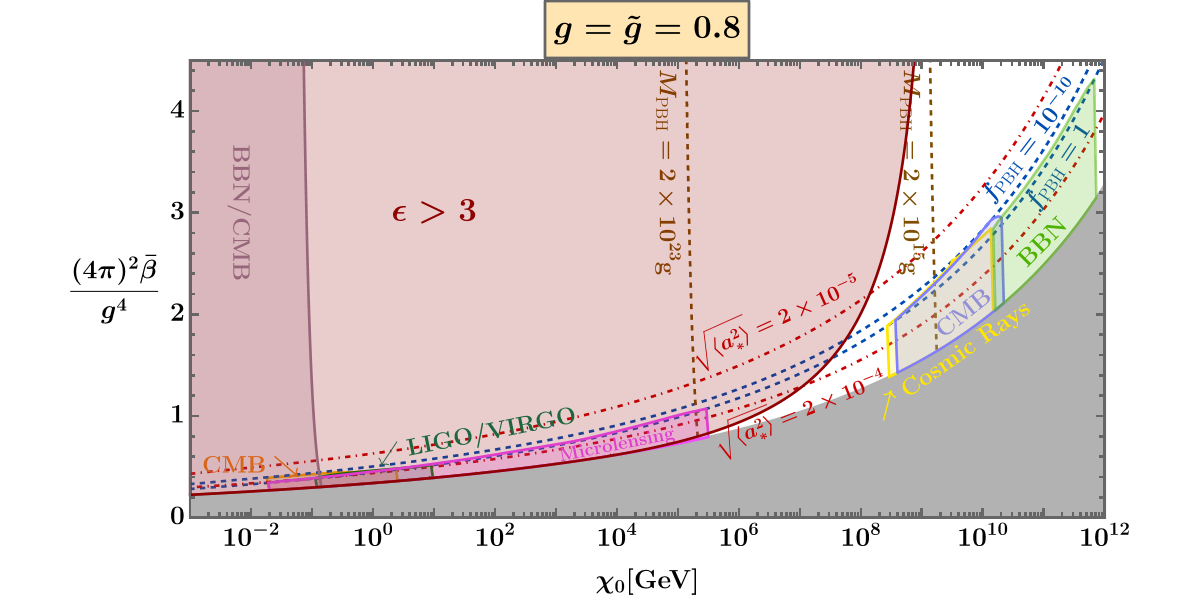}
	
	\vspace{-0.45cm}
	
		\includegraphics[width=0.77\linewidth]{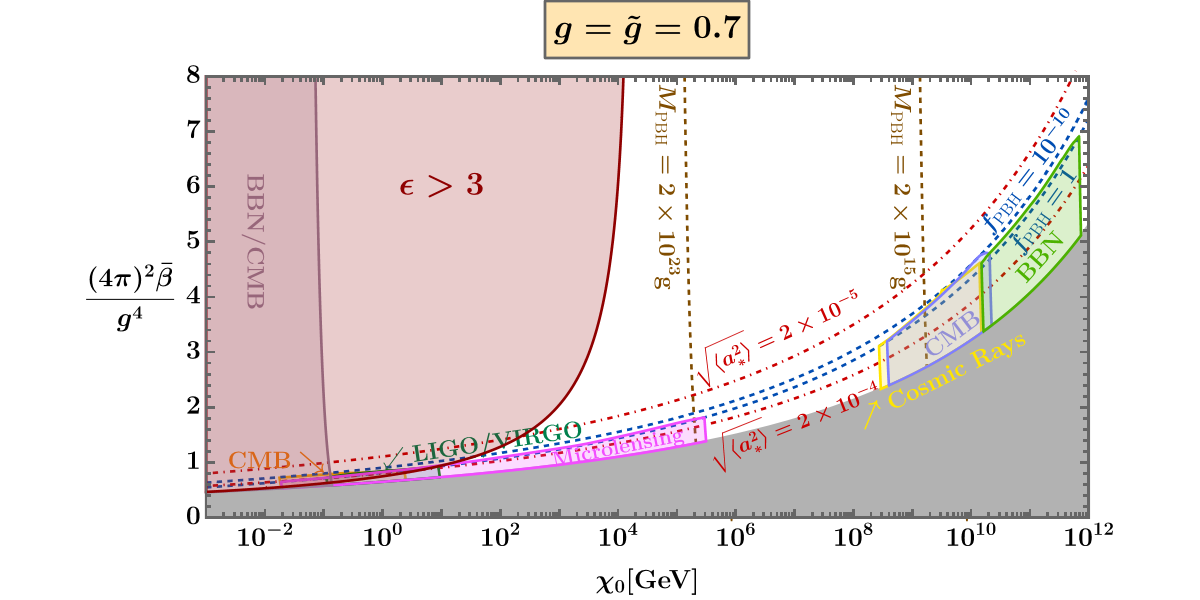} 
		\caption{\em $f_{\rm PBH}$, $M_{\rm PBH}$ and  $\sqrt{\langle a _*^{1/2}\rangle}$ calculated using the improved supercool expansion where the effective potential is approximated as in~(\ref{improLO}) for fixed values of $g$ and $\tilde{g} = g$.}
	\label{fig:ychi0807}
\end{figure}

		\begin{figure}[ht!]
		
		\includegraphics[width=0.75\linewidth]{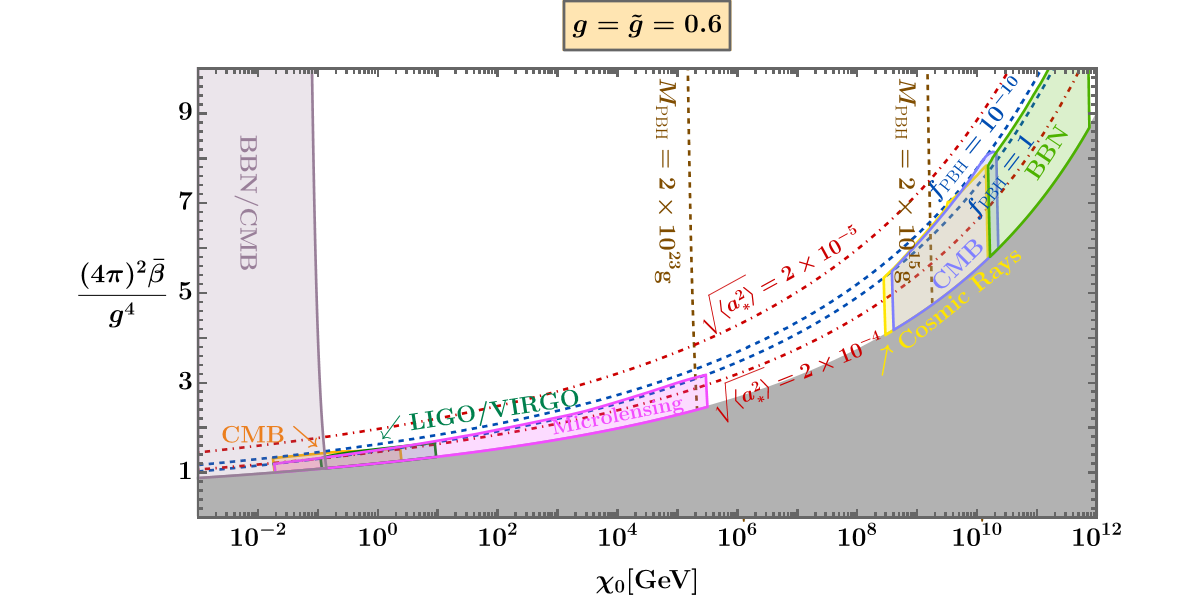}
		
		\vspace{-0.45cm}
		
		\includegraphics[width=0.75\linewidth]{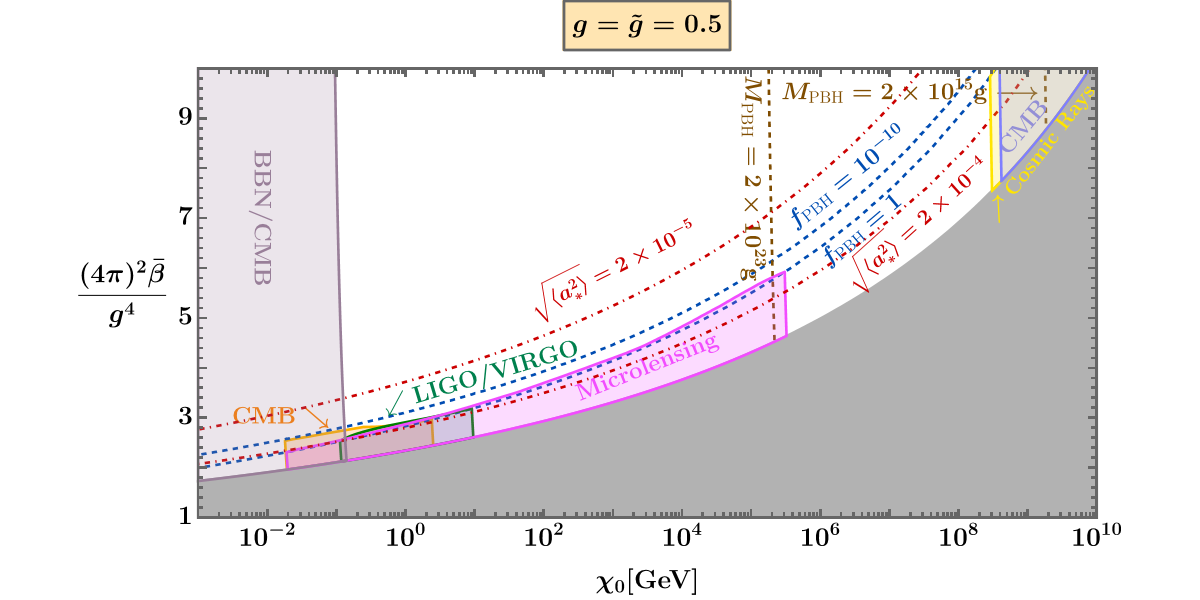}
	\caption{\em Like in Fig.~\ref{fig:ychi0807} but for smaller values of $g$.}
	\label{fig:ychi0807p}
\end{figure}

\begin{figure}[ht!]
		\includegraphics[width=0.36\linewidth]{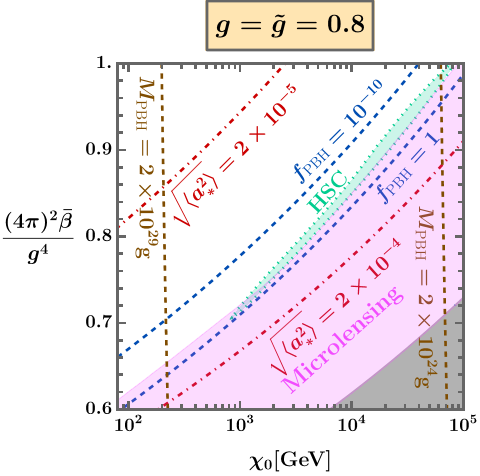} \hspace{0.4cm}
		\includegraphics[width=0.36\linewidth]{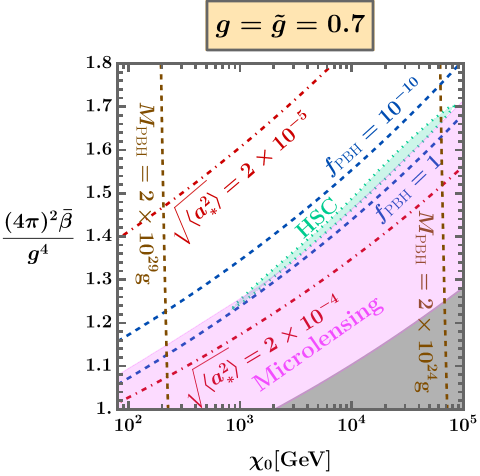} 
	\\
	\vspace{-0.5cm}	
	\includegraphics[width=0.36\linewidth]{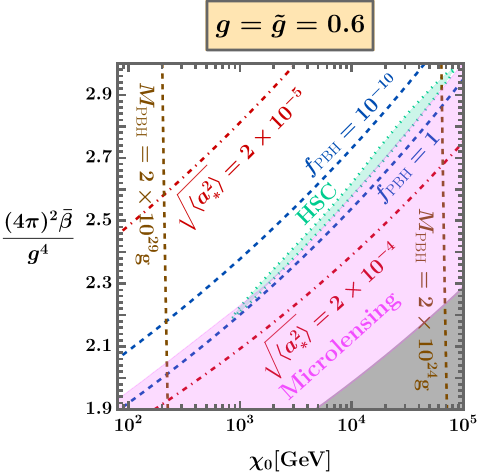} \hspace{0.5cm}
		\includegraphics[width=0.36\linewidth]{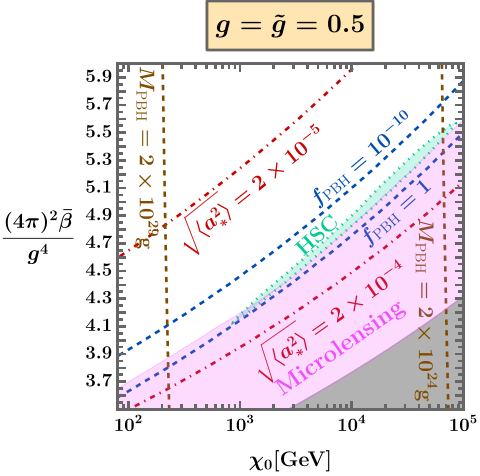}
\\
		\includegraphics[width=0.36\linewidth]{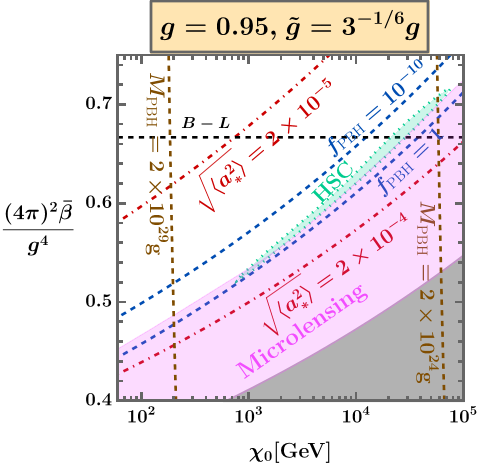} 
	\caption{\em Like in Fig.~\ref{fig:ychi0807} and~\ref{fig:ychi0807p}, but zooming in the best fit regions of  the anomalous microlensing events in HSC 
	data (dotted regions). In the plot in the bottom, $\tilde g/g$ and the  horizontal dashed black line correspond to the values predicted by the model with the $U(1)_{B-L}$ gauge symmetry in~\cite{Salvio:2023ext}.}
		\label{fig:gOGLEHSC}
\end{figure}

In the following plots we highlight the parameter space that allow for a significant PBH production between $f_{\rm PBH}=1$ and $f_{\rm PBH}=10^{-10}$. Also, we represent in {\it dark gray} the region of the parameter space where there is no solution of the equation determining  $T_n$. We also considered the constraints on the PBH abundance:  the accretion constraints due to the CMB~\cite{Serpico:2020ehh}, the merging constraints from LIGO/VIRGO data~\cite{DeLuca:2020qqa}, microlensing constraints (HSC~\cite{Wilkinson:2001vv,Smyth:2019whb}, Kepler~\cite{Griest:2013aaa,Griest:2013esa}, OGLE\footnote{See also~\cite{Mroz:2024wia}.}~\cite{Mroz:2017mvf,Niikura:2019kqi}, EROS~\cite{EROS-2:2006ryy}, MACHO~\cite{Macho:2000nvd}, Icarus~\cite{Oguri:2017ock}, SNe~\cite{Zumalacarregui:2017qqd}), cosmic ray constraints (\hspace{-0.03cm}\cite{Carr:2016hva},~\cite{Carr:2009jm},~\cite{Boudaud:2018hqb}), BBN constraints~\cite{Carr:2009jm} and BBN/CMB bounds on the reheating temperature~\cite{deSalas:2015glj,Hasegawa:2019jsa,Allahverdi:2020bys}.  In the plots we  represent these constraints as follows:  {\it Light Gray}: BBN/CMB constraint requiring that the reheating temperature  be smaller than the neutrino decoupling temperature $\sim 5$ MeV; {\it Orange}: CMB accretion constraint; {\it Dark Green}: merging constraint, {\it Pink}: microlensing constraints, {\it Blue}: CMB evaporation constraint, {\it Yellow}: cosmic ray constraints, {\it Light Green}: BBN constraint. 
The observation or search for GWs necessarily produced by these PTs lead to further constraints; these have been studied extensively and in a fully  model-independent way in Ref.~\cite{Salvio:2023qgb}.

In Figs.~\ref{fig:plote<<1}, ~\ref{fig:ychi0807SE}
 and~\ref{fig:ychi0807SEp} we use the supercool expansion at LO where the effective potential is given by~(\ref{Vlo}). Note that in the red region (where $\epsilon> 1$) the supercool expansion does not necessarily break down: for example, as discussed at the beginning of this section, it can still be valid if there is a sufficiently large number $N$ of degrees of freedom with dominant couplings with $\chi$.

In Figs.~\ref{fig:fixedchi},~\ref{fig:ychi0807} 
and~\ref{fig:ychi0807p} we use the improved supercool expansion where the effective potential is approximated as in~(\ref{improLO}).
In the same approximation we also show a zoom around the electroweak (EW) scale in Fig.~\ref{fig:gOGLEHSC}. There we can show the best fit regions of  the anomalous microlensing events in HSC 
data~\cite{Niikura:2017zjd,Niikura:2019kqi,Sugiyama:2021xqg}.

It is then natural to ask whether there is a specific RSB model able to account for at least some of these anomalous events. 
Ref.~\cite{Salvio:2023ext} constructed\footnote{See also Ref.~\cite{Baldes:2023rqv} for a subsequent related study.} a phenomenological completion of the SM with three right-handed neutrinos (featuring Majorana masses below the EW scale) and  gauged $B-L$. In this model both the EW and $B-L$ gauge symmetries undergo RSB. The gauge group of this model is the SM one times the extra $U(1)$ corresponding to $B-L$, i.e.~$U(1)_{B-L}$, and the field content is the SM one plus the $B-L$ gauge field and a complex scalar.
 In the plot on the bottom of Fig.~\ref{fig:gOGLEHSC} we show that this model fits the HSC data anomaly for $\chi_0\sim2\times 10^4$ GeV and $g=0.95$.
 
 We see that the RSB PTs generically produce PBHs very efficiently without the need of fine tuning. On the contrary, there is a significant parameter space where the PBHs produced in this way have observable effects and a region where they are even overproduced $f_{\rm PBH}>1$. This leads to useful model-independent bounds on the parameter space of theories with RSB.
 
 Regarding the last point, it should be kept in mind, however, that for $M_{\rm PBH}\lesssim 10^{15}$\,g, which corresponds to $\chi_0$ roughly above the $10^9$\,GeV scale,  the PBHs are already evaporated today\footnote{See~\cite{Carr:2020gox} for a review on constraints on PBHs.}. As a result, the $f_{\rm PBH}=1$  line really constrains the parameter space only for\footnote{It is interesting to note that in models with RSB of the Peccei-Quinn symmetry (leading to the QCD axion), see~\cite{DelleRose:2019pgi,VonHarling:2019rgb,Ghoshal:2020vud}, the DM does not receive contribution from the PBHs:  in that case $\chi_0\gtrsim  10^9$\,GeV because of astrophysical observations. Thus in those models DM can be entirely due to the QCD axion.}  $\chi_0\lesssim 10^9$\,GeV.  Nevertheless, from Figs.~\ref{fig:plote<<1}
-\ref{fig:ychi0807p}  we see that for $\chi_0\gtrsim 10^9$\,GeV there are other constraints to keep in mind (the CMB, BBN and cosmic ray  ones in blue, light green and yellow) that are always stronger than the $f_{\rm PBH}=1$  line for those values of $\chi_0$. The parameter space with $M_{\rm PBH}\lesssim 10^{8}$\,g, which corresponds to $\chi_0$  roughly above the $10^{12}$\,GeV scale, remains instead unconstrained.

 We conclude this section with a discussion on the initial spin. Figs.~\ref{fig:plote<<1}
-\ref{fig:gOGLEHSC} show that the RMS value  $\sqrt{\langle a_*^2\rangle}$ is very low. However, there are several mechanisms, which can enhance the PBH spin such as (i) accretion, (ii) close hyperbolic encounters and mergers, (iii) Hawking evaporation in presence of many scalars, etc. Let us briefly elaborate on these mechanisms.

\begin{itemize}
\item \textbf{Accretion.}

There have been several studies which focus on the modification of the PBH mass and spin due to accretion of surrounding matter into the PBH (see~\cite{Ricotti:2007jk,Ricotti:2007au,Berti:2008af,DeLuca:2020bjf} and references therein). The evolution of PBH mass and spin depends on the relative velocity of the PBH and the surrounding matter. As a result the evolution of spin and mass of the PBH takes different directions depending on whether the PBH is isolated (in which case the relative velocity  between the PBH and the surrounding matter  is lower) or there is a binary PBH system (in which case the relative velocity is much larger). Furthermore, since the PBH
might constitute only a fraction of the DM, 
other DM constituents might create a DM halo around the PBH that increases the gravitational potential of the PBH. This renders the accretion more efficient. Note that the plasma-driven superradiant instability may decrease the spin  of the PBH as well, however, for the realistic cases in consideration that effect is negligible.

When the surrounding matter has negligible angular momentum (spherical accretion), the angular momentum of the PBH does not increase, but the mass does, which reduces $a_*$. However, if there is an accretion disk around the PBH, then the non-spherical accretion can become efficient and increase the PBH spin. In Ref.~\cite{DeLuca:2020bjf}, two models have been considered, (I) where the accretion effectively stops around redshift $z\sim 10$ due to structure formation, (II) where the accretion continues even beyond $z\sim 10$. In Model I, PBHs with a final mass (mass after the accretion) below $\approx 30 M_{\odot}$ acquires no spin whereas heavier PBHs can have $a_*\sim 1$. In Model II, even lighter PBHs (PBHs with final mass more than $\approx 10M_{\odot}$) can acquire very high spin for $z\approx 4$. 

\item \textbf{Mergers and Close Hyperbolic Encounters.}

Merger of two non-spinning black holes can give rise to a remnant black hole with significant initial spin. The highest spin of the remnant black hole can be $a_*\approx 0.7$, which occurs in the case of the merger of two equal mass black holes~\cite{Barausse:2009uz} and the value of the spin of the remnant decreases as the mass asymmetry between the two participating black holes increases. Therefore, two PBHs with low initial spin can merge to create a larger black hole with substantial spin.

Furthermore, close hyperbolic encounters is another mechanism that can occur between two PBHs in dense PBH clusters and can induce spins in the PBHs that were initially non-spinning (or very slowly spinning)~\cite{Jaraba:2021ces}. It has been shown that in these encounters, if the two participating PBHs are of the same mass, then the induced spin can go up to $a_*\approx 0.2$. However, if one PBH is much larger than the other, then the induced spin can be considerably larger where the heavier PBH acquires more spin. In this regard, it is to be noted that, for high mass asymmetry, i.e.~if the ratio of the masses of the two black holes is much smaller than unity, the merger occurs. However, before the merger, the participating PBHs will acquire significant spin that can be as high as $0.8$. Furthermore, the acquired spin is larger for higher initial relative velocities and lower impact parameters between the two participating PBHs respectively. Unlike the case of accretion, here the absolute masses of the PBHs do not play a role as the induced spin depends on the ratio of the masses of the two PBHs.

\item \textbf{Hawking Evaporation in the presence of many scalars.}
 
Though a PBH will lose its angular momentum due to evaporation into fermion and vector degrees of freedom, evaporation into scalar degrees of freedom only reduces the PBH mass without changing its angular momentum. Hence, the presence of many scalars can lead to a quick evaporation of a PBH,  which may lead in turn to a PBH spin enhancement.
It is to be noted here that very light PBHs that have already evaporated today may have left traces of their spin. This is because the spectra of the evaporation depends on the spin. 

An example of theories where there are many scalars is the ``string axiverse", in which there can be many (hundreds or even thousands) axion or axion-like particles present in the universe due to  string compactifications~\cite{Arvanitaki:2009fg}. 
In this regard, it has been shown in Ref.~\cite{Calza:2021czr}, that the presence of a few hundred axions can increase the PBH spin from an initial value of $a_*\approx0.01$ to values as high as $a_*\approx0.45$. Therefore, this kind of scenarios can also be very efficient in PBH spin enhancement.
\end{itemize}

\section{Conclusions}\label{Conclusions}

In this work we have performed a detailed analysis of  PBH production in the class of perturbative theories featuring supercooled PTs, those with RSB. A large supercooling is the key ingredient that allowed us to perform a model-independent analysis. We have determined not only the PBH abundance but also the PBH mass and initial spin.

\begin{itemize}
\item A first part of the analysis, Sec.~\ref{A fresh look at the late-blooming mechanism} (and Appendix~\ref{appendix}), provided a fresh look at the late-blooming, a model-independent mechanism for  PBH production in supercooled PTs, assuming the exponential time dependence of $\Gamma$ in Eq.~(\ref{eq:Gamma}). This dependence has often been assumed in the literature. We have then provided a full justification of this assumption in Subsec.~\ref{Time dependence of the false-vacuum decay rate}, exploiting the (improved) supercool expansion of RSB. We find corrections to the exact exponential time dependence, but these corrections are always very small for the time intervals we are interested in. This justification of Eq.~(\ref{eq:Gamma}) allowed us, among other things, to explain why the late-blooming mechanism becomes more effective as $\beta/H_n$ is lowered. Then, in  Sec.~\ref{A fresh look at the late-blooming mechanism} we have also studied the dependence on  $\beta/H_n$ of the time evolution not only of the mass excess of the LP, $\delta$
, but also of the scale factors, the Hubble rates and energy densities both for the background and the LP.
\item Sec.~\ref{General supercooled phase transitions from RSB and PBHs} contains our study of the PBH abundance, mass and initial spin in theories with RSB. The results of this section are ready to use in each model of this sort. We find that there is a large region of the model-independent parameter space  of this class of theories where we have a sizable abundance of PBH, so much that the requirement $f_{\rm PBH}\leq 1$ provides a significant model-independent bound. This bound is, however, present only if the PBHs are heavy enough to survive Hawking radiation until today. We find that the PBH initial spin is generically very small, but, at the end of Sec.~\ref{General supercooled phase transitions from RSB and PBHs} we pointed out  several mechanisms, which can enhance it: e.g.~(i) accretion, (ii) close hyperbolic encounters and mergers and (iii) Hawking evaporation in presence of many scalars.
\item In Subsec.~\ref{PBH abundance, mass and initial PBH spin} we have also found that a concrete model, featuring Majorana masses below the EW scale and  gauged $B-L$ undergoing RSB, can account for the HSC data anomaly in a region if its parameter space. While this anomaly needs to be confirmed, this result shows the testability of the RSB scenario for PBHs and how one can use our model-independent results in specific cases.
\end{itemize}

\appendix

\section{Integration methods for the PBH-production mechanism}\label{appendix}

In this appendix we illustrate the details of our approach to solve the systems of  IDEs in~(\ref{eq:BKG})-(\ref{eq:LHP}). Solving an IDE  is not an easy task; a standard way to proceed is to transform it in a system of ordinary differential equations (ODE) (see~\cite{Flores:2024lng} for the application of this technique to a system of particular relevance for us).

In the following analysis we will find it numerically convenient to define time such that
\be \gamma\equiv\frac{t_{\rm eq} \sqrt{\Delta V}}{\bp} =  \frac{\sqrt{3}}{2} \sinh^{-1}(1) =0.76329...,  \label{gamma}\ee 
which corresponds to 
\be t_{\rm eq}=\frac{\sinh^{-1}(1)}{\sqrt{2} H_{\rm eq}}, \label{teqHeq}\ee 
having used 
\be H_{\rm eq}\equiv H(t_{\rm eq}) = \sqrt{\frac{2\Delta V}{3 \bp^2}}. \label{HeqDef} \ee 

Next, besides~(\ref{NoDimT}), we also define the additional dimensionless variable
\be \hat r(\tau,\tau') \equiv \int_{\tau'}^\tau\frac{d\tilde\tau}{a(\tilde\tau)} =\frac{r(t,t')}{t_{\rm eq}} \ee
and the corresponding single-argument dimensionless variable
\be \hat r(\tau) \equiv \hat r(\tau,\tau_{n_i}), \ee
where $\tau_{n_i}\equiv t_{n_i}/t_{n_i}$ is understood, 
such that
\be \hat r(\tau,\tau')=\hat r(\tau)-\hat r(\tau').\ee
Note that by construction $\hat r(\tau_{n_i}) = 0$.
Using $\hat r(\tau)$ and its powers $\hat r^i(\tau)$ we also construct
\be v_i(\tau) \equiv \int_{\tau_{n_i}}^\tau d\tau' \hat\Gamma(\tau')a^3(\tau')\hat r^i(\tau'),\ee
where $\hat \Gamma(\tau) \equiv t_{\rm eq}^4 \Gamma(t)$, so that for $\tau$ around $\tau_n$
\be \hat\Gamma(\tau)\approx \hat H_n^4
\exp\left(\frac{\beta}{H_n} \hat H_n(\tau-\tau_n)\right),  \qquad \hat H_n \equiv H_n t_{\rm eq}. \label{GammaApp}
\ee
Note that by construction $v_i(\tau_{n_i}) = 0$. Eq.~(\ref{GammaApp}) represents a good approximation for $\hat\Gamma$ in supercooled PTs with RSB, as shown in Subsec.~\ref{Time dependence of the false-vacuum decay rate}.

All these definitions allow us to transform our IDEs into two systems of seven ODEs: they both have the form
\begin{equation}
	\label{eq:7ODE}
	\begin{cases}
		a'(\tau) = \frac{\gamma}{\sqrt{3}}a(\tau)\sqrt{\hat{\rho}_R(\tau)+ (\tau)},\\
		\hat{\rho}'_R(\tau)+4\frac{a'(\tau)}{a(\tau)}\hat{\rho}_R(\tau)=-\hat{\rho}'_V(\tau),\\
		\hat{r}'(\tau)=\frac{1}{a(\tau)},\\
		v_i'(\tau)= \hat{\Gamma}(\tau)a^3(\tau)\hat{r}^i(\tau) \qquad (i=0,1,2,3), \\
	\end{cases},
\end{equation}
where a prime denotes a derivative with respect to $\tau$ and $\hat{\rho}_R\equiv \rho_R/\Delta V$, $\hat{\rho}_V\equiv \rho_V/\Delta V = e^{-I}$  with
\begin{equation}
	\label{eq:Itau}
	I(\tau) = \frac{4\pi}{3}[\hat{r}^3(\tau)v_0(\tau)-3\hat{r}^2(\tau)v_1(\tau)+3\hat{r}(\tau)v_2(\tau)-v_3(\tau)].
\end{equation} 
For the background we solve System~(\ref{eq:7ODE}) for $\tau_{n_i}=\tau_c$, namely the value of $\tau$ at the critical temperature, while for the LP we set $\tau_{n_i}=\tau_{n_i}^{\rm PBH}$. 

Let us now discuss the initial conditions for System~(\ref{eq:7ODE}) by considering first the background and then the LP.  
\begin{itemize}
\item For the {\bf background} we actually approximate $\tau_{n_i} = \tau_c$ with $\tau_{\rm eq}=1$.
This is because the integral $I(t,t_{n_i})$ in~(\ref{eq:F}) receives a contribution from  $t'\in[t_c,t_{\rm eq}]$ that is very small compared to that from $t'>t_{\rm eq}$ ($\Gamma$ is negligibly small for times significantly smaller than $t_n$ and $t_{\rm eq}$ is such a time for supercooled PTs). Thus we set the following initial conditions for the background\footnote{The initial value of $a$ is arbitrary for both the background and the LP as clear from~(\ref{eq:BKG})-(\ref{eq:LHP}).}:
\begin{equation}
	\label{eq:Inbkg}
	\{a(1),\hat{\rho}_R(1),\hat{r}(1),v_i(1)\}	 = \{1,1,0,0\}.
\end{equation}
\item For the {\bf LP} we note that $\tau_{n_i}=\tau_{n_i}^{\rm PBH}>\tau_n>1$, where in the last inequality we used supercooling. Therefore, we first solve the first two equations in~(\ref{eq:7ODE})  for $\tau\in[1, \tau_{n_i}]$ with $\rho_V = \Delta V$, because when $\tau<\tau_{n_i}$ vacuum decay has not yet occurred (by definition of $\tau_{n_i}$). For this first integration we set the initial conditions
\begin{equation}
	\label{eq:InLHP1}
	\{a(1),\hat{\rho}_R(1)\}	= \{1,1\}.
	\end{equation}
	The values $\{a(\tau_{n_i}),\hat\rho_R(\tau_{n_i})\}$ obtained in this first integration are then used as initial conditions for the integration with $\tau\geq \tau_{n_i}$, together with $\{\hat{r}(\tau_{n_i}),v_i(\tau_{n_i})\} = \{0,0\}$, which always hold by construction.
\end{itemize}

 	Note that the only free parameters in the integration are $\beta$ and $\alpha$, which determine $\tau_n$.

 \vspace{1cm}
\footnotesize
\begin{multicols}{2}

\end{multicols}

\end{document}